\documentclass[pre,longbibliography,tighten]{revtex4-1}
\usepackage[utf8]{inputenc}
\usepackage{natbib}
\usepackage{amssymb,amsmath}
\usepackage{epsfig}
\usepackage{bm}
\usepackage{color}

\usepackage{graphicx}

\usepackage{color}

\newcommand\beq{\begin{equation}}
\newcommand\eeq{\end{equation}}
\newcommand\beqa{\begin{eqnarray}}
\newcommand\eeqa{\end{eqnarray}}
\newcommand{\dd}{\text{d}}

\newcommand{\al}{\alpha}

\begin{document}

\title{Mobility and diffusion of intruders in granular suspensions. Einstein relation}

\author{Rub\'en G\'omez Gonz\'alez\footnote[1]{Electronic address: ruben@unex.es}}
\affiliation{Departamento de F\'{\i}sica,
Universidad de Extremadura, E-06006 Badajoz, Spain}
\author{Vicente Garz\'o\footnote[1]{Electronic address: vicenteg@unex.es;
URL: https://fisteor.cms.unex.es/investigadores/vicente-garzo-puertos/}}
\affiliation{Departamento de F\'{\i}sica and Instituto de Computaci\'on Cient\'{\i}fica Avanzada (ICCAEx), Universidad de Extremadura, E-06006 Badajoz, Spain}

\begin{abstract}

The Enskog kinetic equation is considered to determine the diffusion $D$ and mobility $\lambda$ transport coefficients of intruders immersed in a granular gas of inelastic hard spheres (grains). Intruders and grains are in contact with a thermal bath, which plays the role of a background gas. As usual, the influence of the latter on the dynamics of intruders and grains is accounted for via a viscous drag force plus a stochastic Langevin-like term proportional to the background temperature $T_\text{b}$. In this case, the starting kinetic equations are the Enskog and Enskog--Lorentz equations for grains and intruders, respectively, with the addition of Fokker--Planck terms to each one of the above equations. The transport coefficients $\lambda$ and $D$ are determined by solving the Enskog--Lorentz kinetic equation by means of the Chapman--Enskog method adapted to inelastic collisions.
As for elastic collisions, both transport coefficients are given in terms of the solutions of two integral equations which are approximately solved up to the second order in a Sonine polynomial expansion. Theoretical results are compared against numerical solutions of the inelastic Enskog equation by means of the direct simulation Monte Carlo (DSMC) method. Good agreement between theory and simulations is in general found, especially in the case of the second Sonine approximation. The knowledge of the coefficients $\lambda$ and $D$ allow us to assess the departure of the (conventional) Einstein relation $\epsilon=D/(T_{\text{b}}\lambda)$ from 1. As expected from previous results for driven granular gases, it is shown that when the bath temperature $T_\text{b}$ is replaced by the intruder temperature $T_0$ in the Einstein relation, the origin of the deviation of $\epsilon$ from 1 is only due to the non-Maxwellian behavior of reference state of intruders (measured by the cumulant $c_0$). Since the magnitude of $c_0$ is in general very small, deviations of the (modified) Einstein relation $\epsilon_0=D/(T_0\lambda)$ from 1 cannot be detected in computer simulations of dilute granular gases. This conclusion agrees well with previous computer simulation results.  
\end{abstract}

\draft
\date{\today}
\maketitle

\section{Introduction}
\label{sec1}

An interesting challenging problem in statistical mechanics is the generalization of the fluctuation-response relation to non-equilibrium situations \cite{MPRV08}. This problem has received considerable attention in the past few years by many researchers which have tried to carry on such extension not only by means of theoretical tools but also employing computer simulations. Among the different systems which are inherently out of equilibrium, granular matter can be considered as a good candidate to analyze this problem. It is well known that when granular matter is externally excited (rapid flow conditions), the motion of grains resembles the chaotic motion of atoms or molecules in a conventional molecular fluid. However, given that the size of grains is mesoscopic (of the order of 1$\mu$m, for instance), their interactions are inelastic and hence the total energy of the system decreases with time. To keep it in rapid conditions, one has to inject energy into the system to compensate for the energy dissipated by collisions and hence a non-equilibrium steady state (NESS) is reached. For this sort of systems, the fluctuation-response theorem has been proposed in terms of an effective temperature, which is clearly different from the environmental temperature \cite{PBL02,BLP04,SL04,SBL06}.

It is quite apparent that an analysis of the validity of the fluctuation-response theorem requires to know the complete dependence of the response and correlation functions on frequency $\omega$ \cite{M89}. Since this is in fact a quite difficult problem, in order to gain some insight into the general problem one usually considers the limit of small frequencies ($\omega \to 0$). In this limiting case, the classical relation between the diffusion coefficient $D$ (autocorrelation function) and the mobility coefficient $\lambda$ (linear response) is known as the Einstein relation \cite{M89}.

In the case of granular gases, fluctuation response relations have been derived \cite{DG01,DB02} with respect to the so-called homogeneous cooling state (i.e., a state whose fate is a thermal death). In this situation, it has been proven that the response to an external force on an intruder (or impurity) particle violates the usual Einstein relation between the diffusion and mobility coefficients. There are three distinct origins for the violation of the Einstein relation: the deviation of the homogeneous cooling state from the Gibbs state (non-Gaussian distribution functions for the intruder and particles of the granular gas), the cooling of the reference state (yielding a different time-dependence for $D$ and $\lambda$), and energy non-equipartition (leading to different kinetic temperatures between the intruder and gas particles). A different approach widely employed in kinetic theory and computer simulations consists in considering \emph{driven} granular gases where the system is heated by an external force (or \emph{thermostat}) that compensate for the energy lost by collisions. This was the situation studied in Ref.\ \cite{BLP04} by computer simulations; it was shown the validity of the Einstein relation in NESS when the temperature of the bath is replaced by the temperature of the intruder $T_0$ ($\epsilon_0\equiv D/T_0 \lambda=1$). This conclusion also agrees with the results derived from an exactly solvable model for driven dissipative systems \cite{SL04}.

Needless to say, thermostats are introduced to mimic the effects produced by bulk driving as in air-fluidized beds, for instance \cite{SGS05,AD06}. However, unfortunately in most cases it is not clear the relationship between the results derived in driven (thermostated) granular gases and those obtained in real experiments. A more realistic example of thermostated granular systems consists of a set of solid particles surrounded by a gas of molecular particles. This provides a suitable starting point to model the behavior of granular suspensions. When the dynamics of grains is essentially ruled by their collisions, the tools of the classical kinetic theory (conveniently adapted to inelastic collisions) can be a reliable way for describing this type of granular flows \cite{S20}. However, due to the technical difficulties embodied in the study of two or more phases, a coarse-grained approach is generally adopted. In this description, the influence of the background gas on grains is usually incorporated in the kinetic equation through a fluid-solid interaction force. Usually, the gas-phase effects on the solid particles is described by the addition of a Fokker--Planck term (drag force term plus stochastic Langevin-like term) in the kinetic equation \cite{GTSH12}. In fact, this way of driving the granular gas has been employed in computer simulations \cite{BLP04}. It is important to stress that this suspension model can be also derived in a more rigorous way by explicit consideration of the (elastic) collision between grains and particles of the molecular gas. In this discrete description, the above collisions are accounted for via the Boltzmann-Lorentz collision operator \cite{RL77}. In the limit where the grains are much heavier than the molecular gas particles, the Boltzmann-Lorentz operator reduces to the Fokker--Planck operator and the results for transport properties derived from the collision model \cite{GG22} agree with those obtained from the coarse-grained approach \cite{GGG19a}.      

The objective of this paper is to determine the diffusion $D$ and mobility $\lambda$ transport coefficients in a granular suspension. For moderate densities, our starting point are the (nonlinear) Enskog and the (linear) Enskog--Lorentz kinetic equations for the granular gas and the intruders, respectively, with the addition of Fokker--Planck operators to each one of these kinetic equations. The interaction between the grains and intruders with the interstitial gas is through two different drift coefficients $\gamma$ and $\gamma_0$, respectively. To first order in both the concentration gradient and the external field, the Enskog--Lorentz equation is solved by means of the Chapman--Enskog method \cite{CC70} adapted to dissipative dynamics. As for elastic collisions, the coefficients $D$ and $\lambda$ are given in terms of a set of coupled linear integral equations which are approximately solved by considering the second Sonine approximation (i.e., the second order truncation of the Sonine polynomial expansion of the velocity distribution of intruders). As occurs in driven granular gases \cite{G04,G08}, our results show that the deviations of the \emph{modified} Einstein relation $\epsilon_0$ from 1 are only due to the very small departure of the reference state (zeroth-order distribution of intruders) from the Maxwell--Boltzmann distribution. This departure is measured by the kurtosis (or fourth-degree cumulant) $c_0$. Since in general the magnitude of $c_0$ in granular suspensions is much smaller that the one obtained in freely cooling systems \cite{DG01} and/or in driven granular gases \cite{G04,G08}, one may conclude that the verification of the modified Einstein relation is much more accurate in gas-solid flows than in dry granular gases. This is likely one of most relevant conclusions of the present work.   

It is important to remark that our results are based on the Enskog equation. This equation is an extension of the Boltzmann equation (which holds for very dilute gases) to moderate densities. In this regime of densities, although spatial correlations are accounted for via the pair correlation function in this kinetic equation, velocity correlations between the particles which are about to collide are neglected (molecular chaos assumption) as in the Boltzmann description. This is the main limitation of the Enskog equation. In this particular context, it's worth highlighting the computer simulation results obtained by Puglisi \emph{et al.} \cite{PBV07}. These results demonstrate that the departure from the Einstein relation primarily arises from spatial and velocity correlations that emerge with increasing density, rather than the non-Gaussian corrections to the distribution function. This conclusion has been also confirmed by experimental evidence \cite{GPSV14} involving the Brownian motion of a rotating intruder immersed in a vibro-fluidized granular medium. It is shown that Einstein’s relation holds in the dilute regime while it is violated for high packing fraction; this violation cannot be explained in terms of effective temperatures. On the other hand, given that the spatial and velocity correlations are present for densities and inelasticities at which the Enskog equation does not presumably apply, the conclusions reached in Refs.\ \cite{PBV07} and \cite{GPSV14} are not in conflict with those derived here.

The plan of the paper is as follows. In section \ref{sec2}, we describe the problem we are interested in. The steady homogeneous state of the intruders plus granular gas in contact with a thermal bath is studied in section \ref{sec3}. As expected, the intruder's temperature $T_0$ differs from that of the granular gas $T$ and so, there is a breakdown of energy equipartition. In section \ref{sec4}, the Chapman--Enskog method is applied to solve the Enskog--Lorentz kinetic equation to first order in the concentration gradient and the external field. Some technical details concerning the calculations of the paper are provided in the Appendices \ref{appA} and \ref{appB}. The theoretical results for $D$ and $\lambda$ are compared with Monte Carlo simulation results showing an excellent agreement, especially in the case of the second-Sonine solution.
The knowledge of $T$, $T_0$, $D$, and $\lambda$ allows us to compute the conventional $\epsilon$ (defined in terms of the bath temperature $T_\text{b}$) and modified $\epsilon_0$ (defined in terms of the intruder temperature $T_0$) Einstein relations. While $\epsilon_0\simeq 1$, $\epsilon$ clearly differs from 1, showing that the violation of the conventional Einstein relation in granular suspensions can be significant. We close the paper with some concluding remarks.

\section{Description of the problem}
\label{sec2}

\subsection{Granular gas}
\label{subsec1}
We consider a granular gas of inelastic hard spheres of mass $m$ and diameter $\sigma$. The solid particles are immersed in a gas of viscosity $\eta_g$. Spheres (grains) are assumed to be completely smooth so that, inelasticity of collisions is only characterized by the constant (positive) coefficient of normal restitution $\al\leqslant 1$. When the suspensions are dominated by collisions (which are assumed to be nearly instantaneous) \cite{S20}, a coarse-grained description can be adopted to account for the influence of the gas on the dynamics of solid particles. In this approach, the effect of gas-phase on grains
is usually incorporated in the starting kinetic equation  by means of a fluid--solid interaction force \cite{K90,G94,J00}. Some models for granular suspensions \cite{LMJ91,TK95,SMTK96,WZLH09,PS12,H13,WGZS14,SA17,ASG19,SA20} only take into account the Stokes linear drag force law (which attempts to mimic the friction of grains with the interstitial gas) for gas-solid interactions. On the other hand, some works \cite{TGHFS10} have shown that the drag force term does not correctly capture the particle acceleration-velocity correlation observed in direct numerical simulations \cite{TS14}. For this reason, an additional Langevin-like term is included in the effective fluid--solid force. This stochastic term models the additional effects of neighboring particles via the stochastic increment of a Wiener process \cite{GTSH12}. In addition, this term (which randomly kicks the particles between collisions) takes also into account the energy gained by the solid particles due to their interaction with the background gas. 

Thus, according to the above coarse-grained description, for moderate densities and assuming that the granular gas is in a \emph{steady} homogeneous state, the one-particle velocity distribution function $f(\mathbf{v}, t)$ of the granular gas verifies the nonlinear Enskog equation \cite{G19}
\beq
\label{2.1}
-\gamma\frac{\partial}{\partial\mathbf{v}}\cdot\mathbf{v}f-\frac{\gamma T_{\text{b}}}{m}\frac{\partial^2 f}{\partial v^2}=J[\mathbf{v}|f,f],
\eeq
where the Enskog collision operator is
\beq
\label{2.2}
J\left[\mathbf{v}_1|f,f\right]=\chi \sigma^{d-1}\int d\mathbf{v}_2\int \dd\widehat{\boldsymbol{\sigma}}\Theta\left(\widehat{\boldsymbol{\sigma}}\cdot\mathbf{g}_{12}\right)
\left(\widehat{\boldsymbol{\sigma}}\cdot\mathbf{g}_{12}\right)
\Big[\alpha^{-2}f(\mathbf{v}_1'',t)f(\mathbf{v}_2'',t)-f(\mathbf{v}_1,t)f(,\mathbf{v}_2,t)\Big].
\eeq
Here, $\chi$ is the pair correlation function for grain-grain collisions at contact (i.e., when the distance between their centers is $\sigma$), $\widehat{\boldsymbol{\sigma}}$ is a unit vector directed along the line of centers of the colliding spheres, $\Theta$ is the Heaviside step function [$\Theta(x)=1$ for $x> 0$, $\Theta(x)=0$ for $x\leq 0$], and $\mathbf{g}_{12}=\mathbf{v}_1-\mathbf{v}_2$ is the relative velocity of the two colliding spheres. The double primes on the velocities denote the initial values $(\mathbf{v}_1'',\mathbf{v}_2'')$ that yield $(\mathbf{v}_1,\mathbf{v}_2)$ following a binary collision:
\beq
\label{2.3}
\mathbf{v}_1''=\mathbf{v}_1-\frac{1+\alpha^{-1}}{2}\left(\boldsymbol{\widehat{\sigma}}
\cdot\mathbf{g}_{12}\right)\boldsymbol{\widehat{\sigma}}, \quad \mathbf{v}_2'=\mathbf{v}_2+\frac{1+\alpha^{-1}}{2}\left(\boldsymbol{\widehat{\sigma}}
\cdot\mathbf{g}_{12}\right)\boldsymbol{\widehat{\sigma}}.
\eeq

In Eq.\ \eqref{2.1}, $\gamma$ is the drift or friction coefficient (characterizing the interaction between particles of the granular gas and the background gas) and $T_\text{b}$ is the bath temperature. As in previous works \cite{GGG19a,GKG20}, we assume here that $\gamma$ is a scalar quantity proportional to the gas viscosity \cite{KH01}. In the \emph{dilute} limit every particle is only subjected to its respective Stokes drag and so, for hard spheres ($d=3$) the drift coefficient $\gamma$ is defined as
\beq
\label{2.4.0}
\gamma\equiv \gamma_\text{St}=\frac{3 \pi \sigma \eta_g}{m}.
\eeq
Beyond the dilute limit, for moderate densities and low Reynolds numbers, one has the relationship
\beq
\label{2.5}
\gamma=\gamma_\text{St}R(\phi),
\eeq
where $R(\phi)$ is a function of the solid volume fraction
\beq
\label{2.6}
\phi=\frac{\pi^{d/2}}{2^{d-1}d\Gamma \left(\frac{d}{2}\right)}n\sigma^d.
\eeq
The density dependence of the dimensionless function $R$ can be inferred from computer simulations. Specific forms of $R$ will be chosen later for assessing the dependence of the dynamic properties of the system on the parameter space of the problem. On the other hand, it is worthwhile remarking that the results reported in this paper apply regardless of the specific choice of the function $R$.

In the homogeneous state, the only nontrivial balance equation is that of the granular temperature $T$ defined as 
\beq
\label{2.8}
d n T=\int d\mathbf{v}\; m v^2 f(\mathbf{v}),
\eeq
where
\beq
\label{2.9}
n=\int d\mathbf{v}\; f(\mathbf{v})
\eeq
is the number density of solid particles. The balance equation for $T$ can be easily derived by multiplying both sides of Eq.\ \eqref{2.1} by $m v^2$ and integrating over velocity. It is given by
\beq
\label{2.11}
2\gamma \left(T_\text{b}-T \right)=T \zeta,
\eeq
where
\beq
\label{2.12}
\zeta=-\frac{1}{d n T}\int d\mathbf{v}\; m v^2\; J[f,f]
\eeq
is the cooling rate. This quantity gives the rate of change of energy dissipated by collisions. When collisions are elastic ($\al=1$), $\zeta=0$. Since $\zeta$ is a functional of the distribution $f(\mathbf{v})$, it is quite obvious that one needs to know $f$ to determine the cooling rate.

In the case of elastic collisions, Eq.\ \eqref{2.11} leads to the result $T=T_\text{b}$ and the Enskog equation \eqref{2.1} admits the simple Maxwell--Boltzmann solution
\beq
\label{2.13}
f(\mathbf{v})=f_\text{b,M}(\mathbf{v})=n\left(\frac{m}{2\pi T_\text{b}}\right)^{d/2}\exp\left(-\frac{m v^2}{2 T_\text{b}}\right).
\eeq
This result is nothing more than a consequence of the fluctuation-dissipation theorem \cite{K07}. On the other hand, for inelastic collisions ($\al \neq 1$), the exact solution of Eq.\ \eqref{2.1} is not known. However, in the region of thermal velocities, a good approximation can be obtained from an expansion in Sonine polynomials. In the leading order, the distribution $f$ can be written as
\beq
\label{2.14}
f(\mathbf{v}) \to n \pi^{-d/2} v_\text{th}^{-d} e^{-\xi^2}\Bigg\{1+\frac{c}{2}\Bigg[\xi^4-(d+2)\xi^2+\frac{d(d+2)}{4}\Bigg]\Bigg\},
\eeq
where $\boldsymbol{\xi}=\mathbf{v}/v_\text{th}$ and $v_\text{th}=\sqrt{2T/m}$ is a thermal speed. The coefficient $c$ (which measures the deviation of $f$ from its Maxwellian form) is related to the kurtosis of the distribution. Its value has been estimated from the Enskog equation by considering linear terms in $c$ \cite{GGG19a}. Its explicit expression is
\beq
\label{2.15}
c=\frac{16(1-\al)(1-2\al^2)}{73+56d-3\al(35+8d)+30(1-\al)\al^2+\frac{64d(d+2)}{1+\al}\gamma^*},
\eeq
where
\beq
\label{2.16}
\gamma^*=\frac{\gamma}{\nu}=\frac{\sqrt{\pi}}{2^{d} d}\frac{R}{\phi\chi\sqrt{T^*}}.
\eeq
Here, $T^*=T/\mathcal{T}$, $\mathcal{T}=m\sigma^2 \gamma_\text{St}^2$, and we have introduced the effective collision frequency
\beq
\label{2.17}
\nu=\frac{\sqrt{2}\pi^{(d-1)/2}}{\Gamma\left(\frac{d}{2}\right)}n\sigma^{d-1}\chi v_\text{th}.
\eeq
The cooling rate $\zeta$ can be also determined from the Sonine approximation \eqref{2.14} with the result
\beq
\label{2.18}
\zeta=\frac{1-\al^2}{d}\left(1+\frac{3}{16}c\right)\nu.
\eeq
Upon obtaining Eq.\ \eqref{2.18}, nonlinear terms in $c$ have been neglected. 

For practical purposes, it is convenient to write Eq.\ \eqref{2.11} in dimensionless form. In this case, one achieves the equation
\beq
\label{2.19}
2\delta \left(T_b^*-T^*\right)=\zeta^* T^{*3/2},
\eeq
where $T_\text{b}^*=T_\text{b}/\mathcal{T}$, $T^*=T/\mathcal{T}$, $\zeta^*=\zeta/\nu$, and
\beq
\label{2.16.0}
\delta=\frac{\sqrt{\pi}}{2^{d} d}\frac{R}{\phi\chi}.
\eeq
\begin{figure}[h!]
\centering
\includegraphics[width=0.4\textwidth]{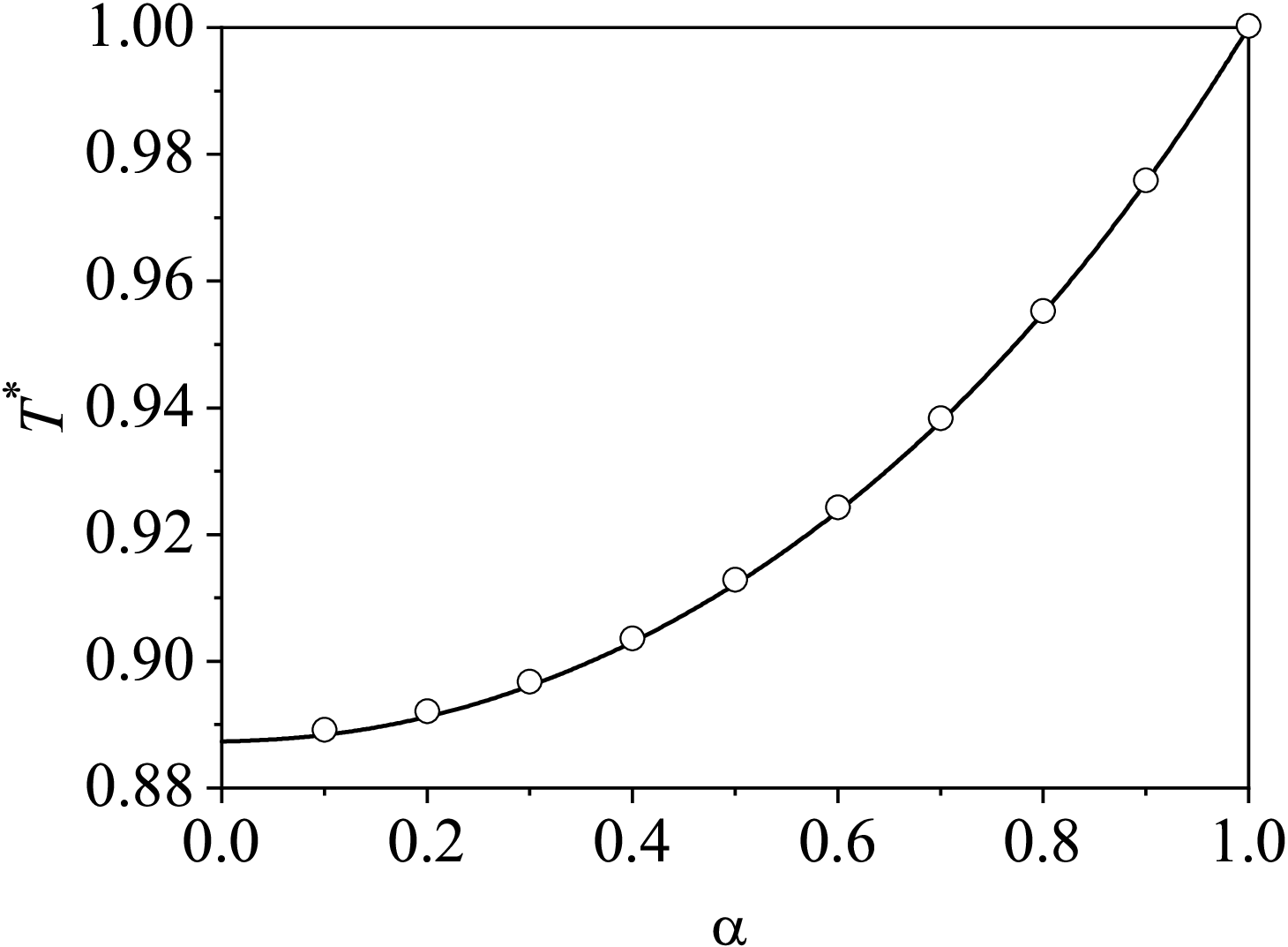}
\caption{Plot of the reduced granular temperature $T^*$ as a function of the coefficient of restitution $\alpha$ for a three-dimensional ($d=3$) system with $T_\text{b}^*=1$ and $\phi=0.1$. The symbols refer to DSMC results.}
\label{fig1}
\end{figure}
\begin{figure}[h!]
\centering
\includegraphics[width=0.4\textwidth]{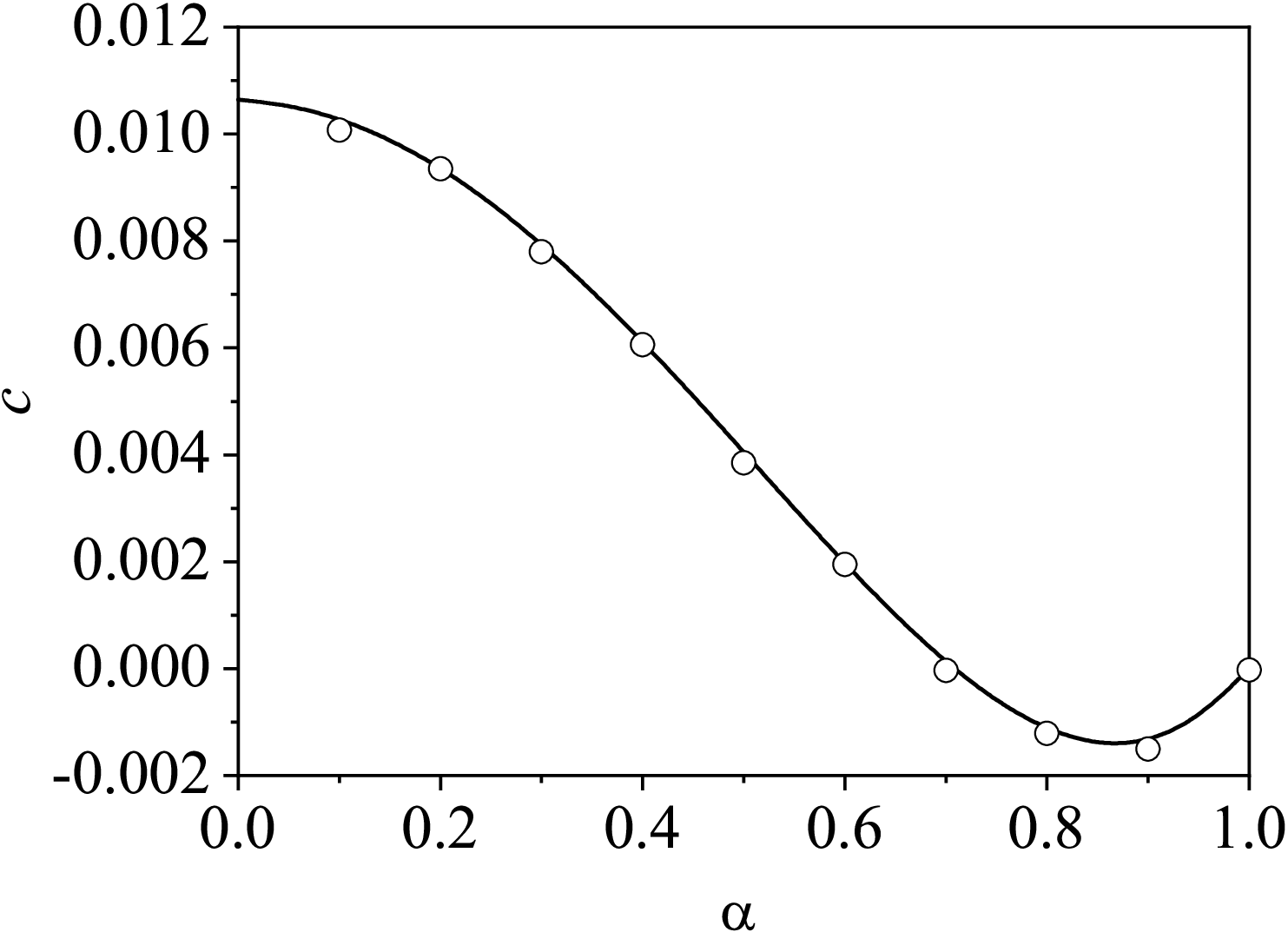}
\caption{Plot of the fourth cumulant $c$ as a function of the coefficient of restitution $\alpha$ for a three-dimensional ($d=3$) system with $T_\text{b}^*=1$ and $\phi=0.1$. The symbols refer to DSMC results.}
\label{fig2}
\end{figure}

If one neglects the kurtosis $c$ (which is in general very small \cite{GGG19a}) in the expression \eqref{2.18} of $\zeta^*$, then
Eq. \eqref{2.19} becomes a cubic equation for the (reduced) temperature $T^*$. In terms of the auxiliary parameter $\varepsilon\equiv \zeta^* \sqrt{T_\text{b}^*}/(2\delta)$, the physical (real) root of the cubic equation \eqref{2.19} can be written as
\beq
\label{2.16.1}
T^*=\frac{\left(\Xi^{1/3}+\Xi^{-1/3}-1\right)^2}{9\varepsilon^2}T_\text{b}^*,
\eeq
where
\beq
\label{2.17.1}
\Xi=\frac{3\sqrt{3}\sqrt{27\varepsilon^4-4\varepsilon^2}+27\varepsilon^2-2}{2}.
\eeq
As expected, for elastic collisions ($\al=1$), $\varepsilon\to 0$ and so, $T^*=T_\text{b}^*$ for any value of $\phi$ and $T_\text{b}^*$. When $\al<1$, $T^*<T_\text{b}^*$ since the granular temperature is smaller than that of the background gas. In the case that the coefficient $c$ is not neglected, Eq.\ \eqref{2.19} is a quartic equation whose physical solution must be numerically determined.

The theoretical results for the (reduced) temperature $T^*$ are compared against DSMC simulations \cite{B94}.  The DSMC simulations are performed following the same steps as described in Refs.\ \cite{GGG23} and \cite{GABG23}.  Notably, modifications have been introduced to the collision stage of the simulation algorithm originally employed by Montanero and Garz\'o \cite{MG02}, with the primary objective of incorporating two key considerations: (i) the tracer concentration of intruder particles and (ii) the impact of the interstitial gas on the dynamic behavior of solid particles. The former adjustment entails the exclusion of intruder-intruder collisions and the preservation of grain velocities after grain-intruder collisions. In contrast, the latter modification exhibits a higher degree of complexity.  For a three-dimensional system ($d=3$), the influence of the interstitial fluid on grains is incorporated by iteratively updating the velocity vector $\mathbf{v}_k$ of each individual grain belonging to species $i$ after every time increment $\delta t$, in accordance with the rule \cite{KG14}:
\beq
\label{DSMC}
\mathbf{v}_k\to e^{-\gamma_i\delta t}\mathbf{v}_k+\left(\frac{6\gamma_i T_\text{b} \delta t}{m_i}\right)^{1/2}\boldsymbol{\varpi}_k.
\eeq
Here, $\boldsymbol{\varpi}_k$ is a random vector of zero mean and unit variance. Equation \eqref{DSMC} converges to the Fokker--Plank operator
when the time step $\delta t$ is much shorter than the mean free time between collisions \cite{KG14}.

Figure \ref{fig1} shows the dependence of the (reduced) granular temperature $T^*$ on the coefficient of restitution $\al$ for a three-dimensional ($d=3$) granular gas with $T_\text{b}^*=1$ and $\phi=0.1$. For $d=3$, a good approximation to the pair correlation function $\chi$ is \cite{CS69}
\beq
\label{2.18.1}
\chi=\frac{1-\frac{1}{2}\phi}{\left(1-\phi\right)^3}.
\eeq
Moreover, for the sake of illustration, simulations for hard spheres systems  \cite{HBK05,BHK07,YS09b} suggest the following form for the function $R(\phi)$:
\beq
\label{2.19.1}
R(\phi)=\frac{10 \phi}{(1-\phi)}+\left(1-\phi\right)^3\left(1+1.5 \sqrt{\phi}\right).
\eeq
The line in Fig.\ \ref{fig1} corresponds to the numerical solution to Eq.\ \eqref{2.19} while the symbols refer to the numerical results obtained from DSMC simulations \cite{B94}. As expected, the energy dissipated by collisions increases with increasing inelasticity and so, the kinetic energy of grains (or equivalently, their reduced temperature $T^*$) decreases. We also observe an excellent agreement between theory and simulations in Fig.\ \ref{fig1} in the complete range of values $\al$. Although not plotted, the curve given by the exact solution \eqref{2.16.1} (i.e., when one neglects $c$) is indistinguishable from the one represented in Fig.\ \ref{fig1} taking into account the value of $c$. This feature can be easily explained by Fig.\ \ref{fig2} where the $\al$ dependence of the fourth cumulant $c$ is plotted for the same system as that of Fig.\ \ref{fig1}. We find that the magnitude of $c$ is very small; much smaller than in the case of (dry) granular gases \cite{NE98}. Moreover, the cumulant $c$ exhibits a non-monotonic dependence on $\al$ since it decreases first as increasing inelasticity, reaches a minimum and then increases with decreasing $\al$. As in the case of $T^*$, an excellent agreement between theory and simulations is found.

\subsection{Intruders immersed in a granular gas}

We assume now that a few intruders (of mass $m_0$ and diameter $\sigma_0$) are added to the system. Since the concentration of intruders is negligibly small, one can assume that the state of the granular gas is not disturbed by the presence of the intruders and hence, its distribution function $f(\mathbf{v})$ obeys the Enskog equation \eqref{2.1}. Moreover, one can also neglects collisions among intruders themselves in the kinetic equation of the one-particle velocity distribution function $f_0(\mathbf{r}, \mathbf{v};t)$ of intruders. Thus, in this limiting tracer case, only the intruder-granular gas collisions (which are characterized by the coefficient of restitution $\al_0\neq \al$) will be considered in the above kinetic equation. Intruders also interact with the interstitial fluid through the friction coefficient $\gamma_0$, which is in general different from $\gamma$. Since we are also interested in obtaining the mobility of intruders, we will assume that intruder particles are also subjected to the action of a weak external field $\mathbf{E}$ (e.g., gravity or an electric field). This field only acts on intruders. 

Note that formally the system (intruder plus granular gas) can be regarded as a binary granular suspension where one of the species is present in tracer concentration. For conciseness, in the remainder we will refer to intruders immersed in a granular suspension instead of a binary granular suspension with one tracer species.

Under the above conditions, the one-particle velocity distribution function $f_0(\mathbf{r}, \mathbf{v};t)$ of intruders verifies the Enskog--Lorentz kinetic equation
\beq
\label{2.20}
\frac{\partial f_0}{\partial t}+\mathbf{v}\cdot \nabla f_0+\frac{\mathbf{E}}{m_0}\cdot \frac{\partial}{\partial\mathbf{v}}\cdot f_0
-\gamma_0\frac{\partial}{\partial\mathbf{v}}\cdot\mathbf{v}f_0-\frac{\gamma_0 T_{\text{b}}}{m_0}\frac{\partial^2 f_0}{\partial v^2}=J_0[\mathbf{r}, \mathbf{v}|f_0,f],
\eeq
where the Enskog--Lorentz collision operator $J_0[f_0,f]$ is \cite{G19}
\beqa
\label{2.21}
J_{0}\left[\mathbf{r}_1, \mathbf{v}_1|f_0,f\right]&=&\overline{\sigma}^{d-1}\int \dd\mathbf{v}_2\int \dd\widehat{\boldsymbol{\sigma}}\Theta\left(\widehat{\boldsymbol{\sigma}}\cdot\mathbf{g}_{12}\right)
\left(\widehat{\boldsymbol{\sigma}}\cdot\mathbf{g}_{12}\right)\Big[\alpha_{0}^{-2}\chi_{0}
(\mathbf{r}_1,\mathbf{r}_1-\boldsymbol{\overline{\sigma}})f_0(\mathbf{r}_1,\mathbf{v}_1'',t)
f(\mathbf{v}_2'')\nonumber\\
& & 
-\chi_{0}(\mathbf{r}_1,\mathbf{r}_1+\boldsymbol{\overline{\sigma}})f_0(\mathbf{r}_1,\mathbf{v}_1,t)
f(\mathbf{v}_2)\Big].
\eeqa
Here, $\chi_0$ is the pair correlation function for intruder-granular gas collisions,  $\boldsymbol{\overline{\sigma}}=\overline{\sigma} \widehat{\boldsymbol{\sigma}}$, $\overline{\sigma}=(\sigma+\sigma_0)/2$, and $\widehat{\boldsymbol{\sigma}}$ is the unit vector directed along the line of centers from the sphere of intruder to the sphere of the granular gas at contact. The relationship between the velocities
$(\mathbf{v}_1'',\mathbf{v}_2'')$ and $(\mathbf{v}_1,\mathbf{v}_2)$ is
\beq
\label{2.22}
\mathbf{v}_1''=\mathbf{v}_1-\mu \left(1+\alpha_0^{-1}\right)\left(\boldsymbol{\widehat{\sigma}}
\cdot\mathbf{g}_{12}\right)\boldsymbol{\widehat{\sigma}}, \quad \mathbf{v}_2''=\mathbf{v}_2+\mu_0 \left(1+\alpha_0^{-1}\right)\left(\boldsymbol{\widehat{\sigma}}
\cdot\mathbf{g}_{12}\right)\boldsymbol{\widehat{\sigma}},
\eeq
where 
\beq
\label{2.22.1}
\mu=\frac{m}{m+m_0}, \quad \mu_0=\frac{m_0}{m+m_0}.
\eeq
Equations \eqref{2.22} give the so-called inverse or \emph{restituting} collisions. The so-called \emph{direct} collisions are defined as collisions where the pre-collisional velocities $(\mathbf{v}_1,\mathbf{v}_2)$ yield $(\mathbf{v}_1',\mathbf{v}_2')$ as post-collisional velocities. Inversion of the collision rules \eqref{2.22} give the forms
\beq
\label{2.24}
\mathbf{v}_1'=\mathbf{v}_1-\mu \left(1+\alpha_0\right)\left(\boldsymbol{\widehat{\sigma}}
\cdot\mathbf{g}_{12}\right)\boldsymbol{\widehat{\sigma}}, \quad \mathbf{v}_2'=\mathbf{v}_2+\mu_0 \left(1+\alpha_0\right)\left(\boldsymbol{\widehat{\sigma}}
\cdot\mathbf{g}_{12}\right)\boldsymbol{\widehat{\sigma}}.
\eeq
Moreover, note that upon writing Eq.\ \eqref{2.21} we have accounted for that the granular gas is in a steady homogeneous state.

In accordance with Eq.\ \eqref{2.5}, the friction coefficient $\gamma_0$ for the intruder can be written as
\beq
\label{2.26}
\gamma_0=\gamma_{0,\text{St}}R_0,
\eeq
where for $d=3$,
\beq
\label{2.27}
\gamma_{0,\text{St}}=\frac{3 \pi \sigma_0 \eta_g}{m_0}=\frac{\sigma_0 m}{\sigma m_0}\gamma_{\text{St}}.
\eeq
As in the case of $R(\phi)$, the dependence of the function $R_0$ on the density $\phi$ and the remaining parameters of the system will be taken from the results obtained by computer simulations.

Apart from the granular temperature $T$, it is convenient at a kinetic level to introduce the partial temperature of intruders $T_0$. This quantity measures the mean kinetic energy of intruders. It is defined as
\beq
\label{2.28}
d n_0(\mathbf{r};t) T_0(\mathbf{r};t)=\int d\mathbf{v}\; m_0 v^2 f_0((\mathbf{r}, \mathbf{v}; t),
\eeq
where
\beq
\label{2.29}
n_0(\mathbf{r};t)=\int d\mathbf{v}\; f_0(\mathbf{r}, \mathbf{v}; t)
\eeq
is the number density of intruders. Upon writing Eq.\ \eqref{2.28} we have taken into account that the mean flow velocity of the granular gas vanishes in our problem. It must recalled that $n_0$ is much smaller than its counterpart $n$ for the particles of the granular gas.

\section{Homogeneous steady state for intruders}
\label{sec3}

Before considering the diffusion of intruders due to the presence of a weak concentration gradient $\nabla n_0$ and/or a weak external field $\mathbf{E}$, it is convenient to characterize first the homogeneous steady state of intruders. This is a crucial point since the latter state plays the role of the reference state in the Chapman--Enskog solution to Eq.\ \eqref{2.20}.

In the absence of diffusion (homogeneous steady state), Eq.\ \eqref{2.20} becomes
\beq
\label{3.1}
-\gamma_0\frac{\partial}{\partial\mathbf{v}}\cdot\mathbf{v}f_0-\frac{\gamma_0 T_{\text{b}}}{m_0}\frac{\partial^2 f_0}{\partial v^2}=\chi_0 J_0^\text{B}[f_0,f],
\eeq
where the Boltzmann--Lorentz operator $J_0^\text{B}[f_0,f]$ is  
\beq
\label{3.1.1}
J_0^\text{B}[f_0,f]= \overline{\sigma}^{d-1}\int d\mathbf{v}_2\int \dd\widehat{\boldsymbol{\sigma}}\Theta
\left(\widehat{\boldsymbol{\sigma}}
\cdot\mathbf{g}_{12}\right)
\left(\widehat{\boldsymbol{\sigma}}\cdot\mathbf{g}_{12}\right)
\Big[\alpha_0^{-2}f_0(\mathbf{v}_1'',t)f(\mathbf{v}_2'',t)-f_0(\mathbf{v}_1,t)f(,\mathbf{v}_2,t)\Big].
\eeq
The equation for the (steady) partial temperature $T_0$ can be easily derived from Eq.\ \eqref{3.1} as
\beq
\label{3.2}
2\gamma_0 \left(T_\text{b}-T_0\right)=T_0 \zeta_0,
\eeq
where
\beq
\label{3.3}
\zeta_0=-\frac{\chi_0}{d n_0 T_0}\int d\mathbf{v}\; m_0 v^2\; J_0^\text{B}[f_0,f]
\eeq
is the partial cooling rate characterizing the rate of energy dissipated by intruder-grain collisions. As in the case of the granular gas, for elastic collisions ($\al_0=\al=1$), $\zeta_0=0$, $T_\text{b}=T_0$, and Eq.\ \eqref{3.1} has the exact solution
\beq
\label{3.4}
f_0(\mathbf{v})=n_0 \left(\frac{m_0}{2\pi T_\text{b}}\right)^{d/2} \exp \left(-\frac{m_0 v^2}{2 T_\text{b}}\right).
\eeq
As occurs for the granular gas, for inelastic collisions ($\al_0 \neq 1$) the solution to Eq.\ \eqref{3.1} is not known to date.

A good estimate for the partial temperature $T_0$ can be obtained by considering the leading Sonine approximation to $f_0(\mathbf{v})$ \cite{G19}:
\beq
\label{3.6}
f_0(\mathbf{v}) \to n_0 \pi^{-d/2} \beta^{d/2} v_\text{th}^{-d} e^{-\beta \xi^2}\Bigg\{1+\frac{c_0}{2}\Bigg[\beta^2 \xi^4-(d+2)\beta \xi^2+\frac{d(d+2)}{4}\Bigg]\Bigg\}.
\eeq
Here,
\beq
\label{3.7}
\beta=\frac{m_0 T}{m T_0}
\eeq
is the ratio between the mean square velocities of intruders and grains and 
\beq
\label{3.8}
c_0=\frac{1}{d(d+2)}\frac{m_0^2}{n_0 T_0^2}\int d\mathbf{v}\; v^4 f_0(\mathbf{v})-1
\eeq
is the fourth-degree cumulant $c_0$. The use of the Sonine approximation \eqref{3.6} to $f_0$ allows us to compute the partial cooling rate $\zeta_0$ by substituting \eqref{3.6} into Eq.\ \eqref{3.3} and retaining only linear terms in $c$ and $c_0$. The expression of the (reduced) cooling rate $\zeta_0^*=\zeta_0/\nu$ can be written as
\beq
\label{3.9}
\zeta_0^*=\zeta_{00}+\zeta_{01}c_0+\zeta_{02}c,
\eeq
where the explicit forms of $\zeta_{00}$, $\zeta_{01}$, and $\zeta_{02}$ can be found in the Appendix \ref{appA}.

The cumulant $c_0$ can be determined by multiplying both sides of the Enskog equation \eqref{3.1} by $v^4$ and integrating over $\mathbf{v}$. In dimensionless form, the result is
\beq
\label{3.10}
\gamma_0^*\left(1+c_0-\frac{T_\text{b}^*}{T_0^*}\right)=\Sigma_0,
\eeq
where $T_0^*=T_0/\mathcal{T}$,
\beq
\label{3.11}
\gamma_0^*=\frac{\gamma_0}{\nu}=\frac{\gamma_{0,\text{St}}}{\gamma_{\text{St}}} \frac{R_0}{R}\gamma^*,
\eeq
and
\beq
\label{3.12}
\Sigma_0=\frac{\chi_0}{4d(d+2)}\frac{m_0^2}{n_0 T_0^2 \nu}\int d\mathbf{v}\; v^4 J_0^\text{B}[f_0,f].
\eeq
Retaining only linear terms in $c$ and $c_0$, one has the result 
\beq
\label{3.13}
\Sigma_0=\Sigma_{00}+\Sigma_{01}c_0+\Sigma_{02}c,
\eeq
where the explicit forms of $\Sigma_{00}$, $\Sigma_{01}$, and $\Sigma_{02}$ are provided in the Appendix \ref{appA}. The expression of $c_0$ can be easily obtained when one takes into account Eq.\ \eqref{3.13} in Eq.\ \eqref{3.10}. It is given by
\beq
\label{3.14}
c_0=\frac{\gamma_0^*\left(1-\frac{T_\text{b}^*}{T_0^*}\right)-\Sigma_{00}-\Sigma_{02}c}{\Sigma_{01}-\gamma_0^*}.
\eeq

\begin{figure}[h!]
\centering
\includegraphics[width=0.4\textwidth]{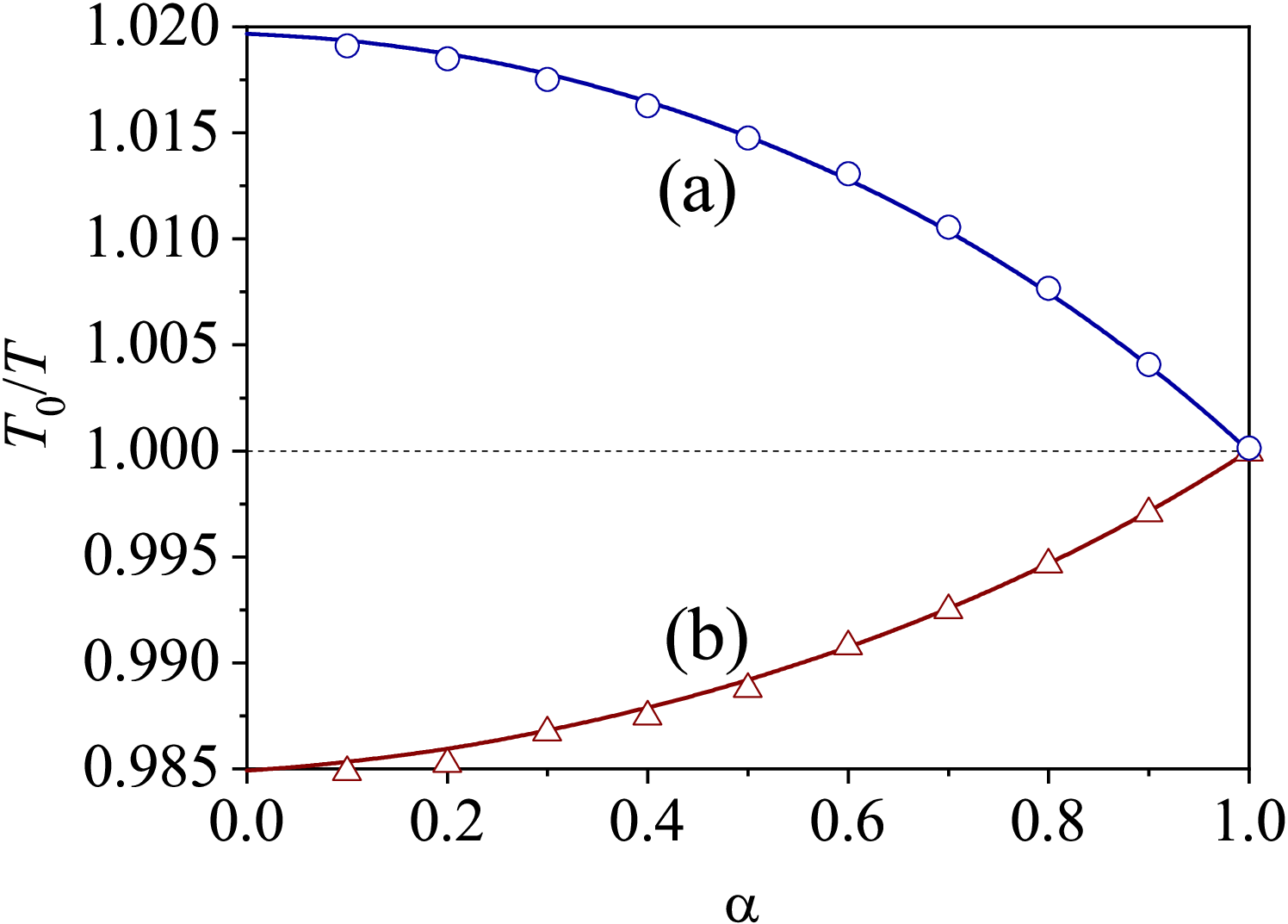}
\caption{Plot of the temperature ratio $T_0/T$ versus the (common) coefficient of restitution $\al$ for a three-dimensional ($d=3$) system with $T_\text{b}^*$, $\phi=0.1$, and two different mixtures: (a) $m_0/m=0.5$ and $\sigma_0/\sigma=1$ (red line and triangles) and (b) $m_0/m=2$ and $\sigma_0/\sigma=1$ (blue line and circles). The symbols refer to the DSMC results.}
\label{fig3}
\end{figure}
\begin{figure}[h!]
\centering
\includegraphics[width=0.4\textwidth]{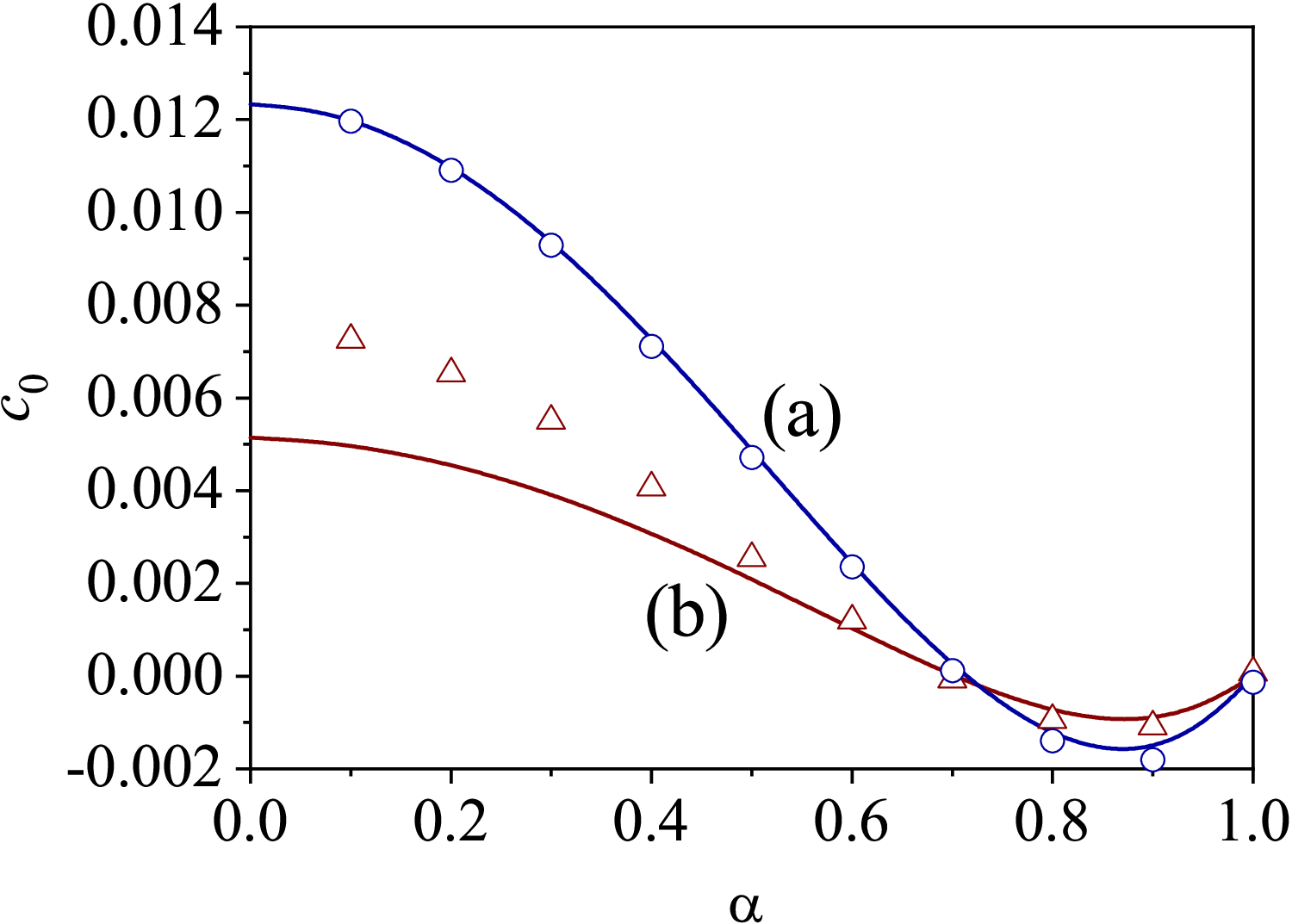}
\caption{Plot of the fourth cumulant of the intruder $c_0$ versus the (common) coefficient of restitution $\al$ for a three-dimensional ($d=3$) system with $T_\text{b}^*$, $\phi=0.1$, and two different mixtures: (a) $m_0/m=0.5$ and $\sigma_0/\sigma=1$ (red line and triangles) and (b) $m_0/m=2$ and $\sigma_0/\sigma=1$ (blue line and circles). The symbols refer to the DSMC results.}
\label{fig4}
\end{figure}

Finally, in dimensionless form, Eq.\ \eqref{3.2} for $T_0^*$ can be written as
\beq
\label{3.15}
2\gamma_0^* \left(T_\text{b}^*-T_0^*\right)=T_0^* \left(\zeta_{00}+\zeta_{01}c_0+\zeta_{02}c\right).
\eeq
Substitution of Eqs.\ \eqref{2.15} and \eqref{3.14} into Eq.\ \eqref{3.15} allows us to determine $T_0^*$ in terms of the parameter space of the system. When intruder and granular gas particles are mechanically equivalent ($m=m_0$, $\sigma=\sigma_0$, and $\al=\al_0$), then $\gamma^*=\gamma_0^*$, $\zeta^*=\zeta_0^*$, and Eq.\ \eqref{3.15} yields $T^*=T_0^*$. This means that energy equipartition applies in the self-diffusion problem. However, in the general case (namely, when collisions are inelastic and intruder and grains are mechanically different), one has to numerically solve Eq.\ \eqref{3.15}. As in the free cooling case \cite{GD99b,MG02,BRM05}, $T_0^*\neq T^*$ and so there is a breakdown of the energy equipartition, as expected.

The dependence of the temperature ratio $T_0/T$ on the (common) coefficient of restitution $\al=\al_0$ is plotted in Fig.\ \ref{fig3} for $d=3$, $T_\text{b}^*=1$, and $\phi=0.1$. Two different mixtures have been considered. For $d=3$, a good approximation for $\chi_0$ is \cite{GH72} 
\beq
\label{3.16}
\chi_0=\frac{1}{1-\phi}+3 \frac{\sigma_0}{\sigma+\sigma_0}\frac{\phi}{(1-\phi)^2}+2 \left(\frac{\sigma_0}{\sigma+\sigma_0}\right)^2 \frac{\phi^2}{(1-\phi)^3}.
\eeq
In addition, in the case of an interstitial fluid with low-Reynolds-number and moderate densities, computer simulations for polydisperse gas-solid flows \cite{HBK05,BHK07,YS09b} estimate $R_0$ as 
\beq
\label{3.17}
R_0=1+\left(R-1\right)\left[a \frac{\sigma_0}{\sigma}+(1-a)\frac{\sigma_0^2}{\sigma^2}\right],
\eeq
where
\beq
\label{3.18}
a(\phi)=1-2.660\phi+9.096\phi^2-11.338\phi^3.
\eeq
Note that for mechanically equivalent particles, one has $R_0=R$, as it should be the case.

In agreement with previous results \cite{GABG23} obtained by neglecting $c$ and $c_0$, Fig.\ \ref{fig3} shows a very tiny impact of the mass and diameter ratios on the temperature ratio $T_0/T$. In fact, this influence is amplified by the scale of the vertical axis. This means that the breakdown of energy equipartititon in granular suspensions is much more modest than in dry granular mixtures \cite{GD99b,MG02} where the ratio $T_0/T$ clearly differs from 1 for both disparate mass and diameter ratios and/or strong inelasticity. At a more qualititative level, we see that $T_0>T$ ($T_0<T$) when the intruder is heavier (lighter) than the particles of the granular gas. This behavior is also present in dry granular mixtures. We find again an excellent agreement between theory and simulations. As a complement of Fig.\ \ref{fig3}, 
Fig.\ \ref{fig4} shows $c_0$ versus $\al$ for the same systems as in Fig.\ \ref{fig3}. As in the case of $c$, the magnitude of the cumulant $c_0$ is very small showing that the deviation of the homogeneous distribution $f_0$ from the Maxwell--Boltzmann distribution is imperceptible in granular suspensions.   
Good agreement between theory and DSMC results is observed, except when $m_0/m=0.5$ for very small values of $\al$ ($\al\lesssim 0.3$). However, these discrepancies are in the order of 2$\%$, which is still lower than in the dry case.

\section{Diffusion and mobility transport coefficients}
\label{sec4}

The objective of this section is to determine the diffusion and mobility transport coefficients of intruders immersed in a granular suspension. As said before, the diffusion process is induced here by the presence of both a weak concentration gradient $\nabla n_0$ and a weak external field $\mathbf{E}$.  The corresponding transport coefficients are obtained by solving the Enskog--Lorentz kinetic equation \eqref{2.20} by means of the Chapman--Enskog method \cite{CC70}. Since the granular gas is in a homogeneous state, $\chi_0$ is constant in the tracer limit and the Enskog--Lorentz operator adopts the simple form  $J_0[f_0,f]=\chi_0 J_0^\text{B}[f_0,f]$.

The intruders may freely exchange momentum and energy in its interaction with the particles of the granular gas; this means that these quantities are not invariants of the Enskog--Lorentz collision operator $J_0[f_0,f]$. Only the number density of intruders $n_0$ is conserved. Its continuity equation can be easily derived from Eq.\ \eqref{2.20} as    
\beq
\label{4.1}
\frac{\partial n_0}{\partial t}=-\nabla \cdot \mathbf{j}_0,
\eeq
where
\beq
\label{4.2}
\mathbf{j}_0(\mathbf{r}; t)= \int d\mathbf{v}\; \mathbf{v}\; f_0(\mathbf{r}, \mathbf{v};t)
\eeq
is the intruder particle flux.

As usual in the Chapman--Enskog method, one assumes the existence of a \emph{normal} solution where all the space and time dependence of $f_0$ only occurs through a functional dependence on the hydrodynamic fields. In this problem, the normal solution to $f_0$ is explicitly generated by expanding this distribution in powers of $\nabla n_0$ and $\mathbf{E}$:
\beq
\label{4.3}
f_0=f_0^{(0)}+\vartheta f_0^{(1)}+\vartheta^2 f_0^{(2)}+\cdots.
\eeq
In Eq.\ \eqref{4.3}, each factor $\vartheta$ corresponds to the implicit factors $\nabla n_0$ and $\mathbf{E}$. Here, only terms to first-order in $\vartheta$ will be considered. The time derivative $\partial_t$ is also expanded as $\partial_t=\partial_t^{(0)}+\vartheta \partial_t^{(1)}+\cdots$, where
\beq
\label{4.4}
\partial_t^{(0)}n_0=0, \quad \partial_t^{(0)}T=2\gamma\left(T_\text{b}-T\right)-\zeta T,
\eeq
\beq
\label{4.5}
\partial_t^{(1)} n_0=-\nabla \cdot \mathbf{j}_0^{(0)}, \quad \partial_t^{(1)}T=0,
\eeq
and
\beq
\label{4.6}
\mathbf{j}_0^{(0)}= \int d\mathbf{v}\; \mathbf{v}\; f_0^{(0)}(\mathbf{v};t).
\eeq

As noted in previous works \cite{GGG19a,GChV13a,GKG20}, although we are interested in computing the diffusion coefficient under steady-state conditions, the presence of the interstitial fluid
introduces the possibility of a local energy unbalance, and
hence, the zeroth-order distribution $f_0^{(0)}$  is not in general a stationary
distribution. This is because for arbitrary small deviations
from the homogeneous steady state the energy gained by
grains due to collisions with the background fluid cannot be
locally compensated with the other cooling terms arising from
the viscous friction and the collisional dissipation. Thus, in
order to get the diffusion and mobility coefficients in the steady state, one has to determine first the \emph{unsteady} integral equation obeying both coefficients and then solve it under the steady-state condition \eqref{2.11}.

The zeroth-order approximation $f_0^{(0)}$ obeys the kinetic equation
\beq
\label{4.7}
\Delta T\frac{\partial f_0^{(0)}}{\partial T}-\gamma_0\frac{\partial}{\partial\mathbf{v}}\cdot\mathbf{v}f_0^{(0)}-\frac{\gamma_0 T_{\text{b}}}{m_0}\frac{\partial^2 f_0^{(0)}}{\partial v^2}=\chi_0 J_0^\text{B}[f_0^{(0)},f],
\eeq
where $\Delta\equiv 2\gamma\left(\frac{T_\text{b}}{T}-1\right)-\zeta$.
Upon deriving Eq.\ \eqref{4.7}, we have accounted for that $f_0^{(0)}$ depends on time through its dependence on temperature $T$. In the steady state ($\Delta=0$), Eq.\ \eqref{4.7} has the same form as Eq.\ \eqref{3.1}. This means that $f_0^{(0)}$ is the solution of Eq.\ \eqref{3.1} but taking into account the local dependence of the density $n_0$. An approximate form to $f_0^{(0)}$ is given by the Sonine approximation \eqref{3.6}. Since $f_0$ is isotropic in velocity, then $\mathbf{j}_0^{(0)}=\mathbf{0}$ and hence $\partial_t^{(1)} n_0=0$.

To first order in $\vartheta$, one achieves the kinetic equation
\beq
\label{4.9}
-\gamma_0\frac{\partial}{\partial\mathbf{v}}\cdot\mathbf{v}f_0^{(1)}-\frac{\gamma_0 T_{\text{b}}}{m_0}\frac{\partial^2 f_0^{(1)}}{\partial v^2}-\chi_0 J_0^\text{B}[f_0^{(1)},f]=-f_0^{(0)}\mathbf{v}\cdot \nabla \ln n_0-\frac{\mathbf{E}}{m_0}\cdot \frac{\partial}{\partial \mathbf{v}}f_0^{(0)}.
\eeq
Upon obtaining Eq.\ \eqref{4.9} we have considered steady conditions ($\Delta=0$) and have taken into account that $\nabla f_0^{(0)}=f_0^{(0)}\nabla \ln n_0$. The solution to Eq.\ \eqref{4.9} can be written as
\beq
\label{4.10}
f_0^{(1)}(\mathbf{v})=\boldsymbol{\mathcal A}(\mathbf{v}) \cdot \nabla \ln n_0+\boldsymbol{\mathcal B}(\mathbf{v})\cdot \mathbf{E},
\eeq
where the coefficients $\boldsymbol{\mathcal A}$ and $\boldsymbol{\mathcal B}$ are functions of the velocity and the hydrodynamic fields. Substitution of Eq.\ \eqref{4.10} into Eq.\ \eqref{4.9} yield the following set of linear integral equation for the unknowns $\boldsymbol{\mathcal A}$ and $\boldsymbol{\mathcal B}$:
\beq
\label{4.11}
-\gamma_0\frac{\partial}{\partial\mathbf{v}}\cdot\mathbf{v}\boldsymbol{\mathcal A}-\frac{\gamma_0 T_{\text{b}}}{m_0}\frac{\partial^2 \boldsymbol{\mathcal A}}{\partial v^2}-\chi_0 J_0^\text{B}[\boldsymbol{\mathcal A},f]=-\mathbf{v} f_0^{(0)},
\eeq
\beq
\label{4.12}
-\gamma_0\frac{\partial}{\partial\mathbf{v}}\cdot\mathbf{v}\boldsymbol{\mathcal B}-\frac{\gamma_0 T_{\text{b}}}{m_0}\frac{\partial^2 \boldsymbol{\mathcal B}}{\partial v^2}-\chi_0 J_0^\text{B}[\boldsymbol{\mathcal B},f]=-\frac{1}{m_0}\frac{\partial}{\partial \mathbf{v}}f_0^{(0)}.
\eeq

In the first order of $\nabla n_0$ and $\mathbf{E}$, the intruder particle flux has the form
\beq
\label{4.13}
\mathbf{j}_0^{(1)}=-D\nabla \ln n_0+\lambda \mathbf{E},
\eeq
where $D$ is the diffusion coefficient and $\lambda$ is the mobility coefficient. Since
\beq
\label{4.14}
\mathbf{j}_0^{(1)}=\int d\mathbf{v}\; \mathbf{v}\; f_0^{(1)}(\mathbf{v}),
\eeq
then, according to Eq.\ \eqref{4.10}, $D$ and $\lambda$ are defined as
\beq
\label{4.15}
D=-\frac{1}{d}\int d\mathbf{v}\; \mathbf{v}\cdot \boldsymbol{\mathcal A}(\mathbf{v}), \quad \lambda=\frac{1}{d}\int d\mathbf{v}\; \mathbf{v}\cdot \boldsymbol{\mathcal B}(\mathbf{v}).
\eeq

For elastic collisions ($\al=\al_0=1$), $T=T_0=T_\text{b}$ and $f_0^{(0)}(\mathbf{v})$ is the local equilibrium distribution \eqref{3.4}. In this case, $\partial f_0^{(0)}/\partial \mathbf{v}=-(m_0 \mathbf{v}/T_\text{b})f_0^{(0)}$ and the integral equations \eqref{4.11} and \eqref{4.12} lead to the identity $\boldsymbol{\mathcal A}=-T_\text{b} \boldsymbol{\mathcal B}$. As a consequence, the conventional Einstein relation is verified, namely,
\beq
\label{4.17}
\epsilon=\frac{D}{T_\text{b} \lambda}=1.
\eeq

On the other hand, for inelastic collisions, $T\neq T_0\neq T_\text{b}$ and hence, the relationship between $D$ and $\lambda$ is no longer simple. There are in principle three different reasons for which the Einstein relation \eqref{4.17} is not verified for granular suspensions. First, when $\al<1$, the granular temperature $T$ is different from the bath temperature $T_\text{b}$ ($T<T_\text{b}$). Second, there is a breakdown of the energy equipartition ($T\neq T_0$) when intruders are mechanically different to the particles of the granular gas. Finally, as a third reason, since $f_0^{(0)}$ is not a Gaussian distribution then, $\partial f_0^{(0)}/\partial \mathbf{v}\neq -(m_0 \mathbf{v}/T_\text{b})f_0^{(0)}$ and hence, $D$ is not proportional to $\lambda$. The first two reasons of discrepancy can be avoided if one replaces the bath temperature $T_\text{b}$ by the intruder particle $T_0$ in the Einstein relation \eqref{4.17}. This change leads to the \emph{modified} Einstein relation
\beq
\label{4.18}
\epsilon_0=\frac{D}{T_{0} \lambda}.
\eeq
The relation \eqref{4.18} was proposed by Barrat \emph{et al.} \cite{BLP04} to extend the Einstein relation \eqref{4.17} to granular gases. 

Note that in particular  if one takes the Maxwellian approximation \eqref{3.4} for $f_0^{(0)}$ with $T_0$ instead of $T_\text{b}$, then $\partial f_0^{(0)}/\partial \mathbf{v}= -(m_0 \mathbf{v}/T_{0})f_0^{(0)}$ and hence $\epsilon_0=1$. Thus, it seems that the only reason for which $\epsilon_0\neq 1$ is due to the absence of the Gibbs state (non-Gaussian behaviour of the distribution $f_0^{(0)}$). Since we have seen that the magnitude of the kurtosis $c_0$  is in general very small for granular suspensions (see for instance Fig.\ \ref{fig4}), one expects the deviations of $\epsilon_0$ from 1 can be quite difficult to detect in computer simulation experiments. In fact, molecular dynamics simulations \cite{BLP04} (for a similar sort of thermostat as the one employed in this paper) did not observe any deviation from the modified Einstein relation ($\epsilon_0=1$) for a wide range of values of the coefficients of restitution and parameters of the mixture. Our objective here to assess the departure of $\epsilon_0$ from 1 in a granular suspension modeled by a stochastic bath with viscous friction.

\subsection{Second Sonine approximation to $D$ and $\lambda$}

It is quite apparent that the transport coefficients $D$ and $\lambda$ are given in terms of the solution of the integral equations \eqref{4.11} and \eqref{4.12}, respectively. These equations can be approximately solved by using a Sonine polynomial expansion. Here, as mentioned in section \ref{sec1}, we determine $D$ and $\lambda$ up to the second Sonine approximation. In this case, $\boldsymbol{\mathcal A}(\mathbf{v})$ and $\boldsymbol{\mathcal B}(\mathbf{v})$ are approximated by
\beq
\label{4.19}
\boldsymbol{\mathcal A}(\mathbf{v})\to -f_{0,\text{M}}(\mathbf{v})\Big[a_1 \mathbf{v}+a_2 \mathbf{S}_0(\mathbf{v})\Big], \quad 
\boldsymbol{\mathcal B}(\mathbf{v})\to -f_{0,\text{M}}(\mathbf{v})\Big[b_1 \mathbf{v}+b_2 \mathbf{S}_0(\mathbf{v})\Big],
\eeq
where
\beq
\label{4.21}
f_{0,\text{M}}(\mathbf{v})=n_0 \left(\frac{m_0}{2\pi T_0}\right)^{d/2} \exp \left(-\frac{m_0 v^2}{2 T_0}\right),
\eeq
and $\mathbf{S}_0(\mathbf{v})$ is the polynomial
\beq
\label{4.22}
\mathbf{S}_0(\mathbf{v})=\Big(\frac{1}{2}m_0 v^2-\frac{d+2}{2}T_0\Big)\mathbf{v}.
\eeq
The Sonine coefficients $a_1$, $b_1$, $a_1$, and $a_2$ are defined as
\begin{equation}
\label{4.23}
\left(
\begin{array}{c}
a_{1}\\
b_1
\end{array}
\right)
=-\frac{m_0}{dn_0T_0}\int d{\bf v}\,{\bf v} \cdot
\left(
\begin{array}{c}
\boldsymbol{\mathcal A}\\
\boldsymbol{\mathcal B}
\end{array}
\right), \quad \left(
\begin{array}{c}
a_{2}\\
b_2
\end{array}
\right)
=-\frac{2}{d(d+2)}\frac{m_0}{n_0T_0^3}\int d{\bf v}\,{\bf S}_0
  \cdot
\left(
\begin{array}{c}
\boldsymbol{\mathcal A}\\
\boldsymbol{\mathcal B}
\end{array}
\right).
\end{equation}
According to Eq.\ \eqref{4.15}, $a_1=m_0D/(n_0 T_0)$ and $b_1=-m_0\lambda/(n_0 T_0)$. The evaluation of the coefficients $a_1$, $b_1$, $a_1$, and $a_2$ is carried out in the Appendix \ref{appB}.

The knowledge of the Sonine coefficients allows us to determine the first- and second-Sonine approximations to the diffusion coefficient $D$ and the mobility coefficient $\lambda$. To write these expressions, it is convenient to introduce the dimensionless coefficients
\beq
\label{4.26}
D^*=\frac{m_0\nu}{T n_0}D, \quad \lambda^*=\frac{m_0\nu}{n_0}\lambda.
\eeq
The second Sonine approximation $D^*[2]$ to $D^*$ can be written as
\beq
\label{4.27}
D^*[2]=\frac{(\nu_4^*+3\gamma_0^*-c_0 \nu_2^*)\tau_0}{(\nu_1^*+\gamma_0^*)(\nu_4^*+3\gamma_0^*)-\nu_2^*\Big[\nu_3^*+2\gamma_0^*
\Big(1-\frac{T_\text{b}^*}{T_0^*}\Big)\Big]},
\eeq
where $\tau_0=T_0/T$ is the temperature ratio. 
The expressions of the (reduced) collision frequencies $\nu_1^*$--$\nu_4^*$ can be found in the Appendix \ref{appB}.  Equation \eqref{4.27} agrees with previous results derived in Ref.\ \cite{GABG23} when one takes $c=c_0=0$. The second-Sonine approximation $\lambda^*[2]$ to $\lambda^*$ is given by
\beq
\label{4.29}
\lambda^*[2]=\frac{\nu_4^*+3\gamma_0^*}{(\nu_1^*+\gamma_0^*)(\nu_4^*+3\gamma_0^*)-\nu_2^*\Big[\nu_3^*+2\gamma_0^*
\Big(1-\frac{T_\text{b}^*}{T_0^*}\Big)\Big]}.
\eeq

\begin{figure}[h!]
\centering
\includegraphics[width=0.4\textwidth]{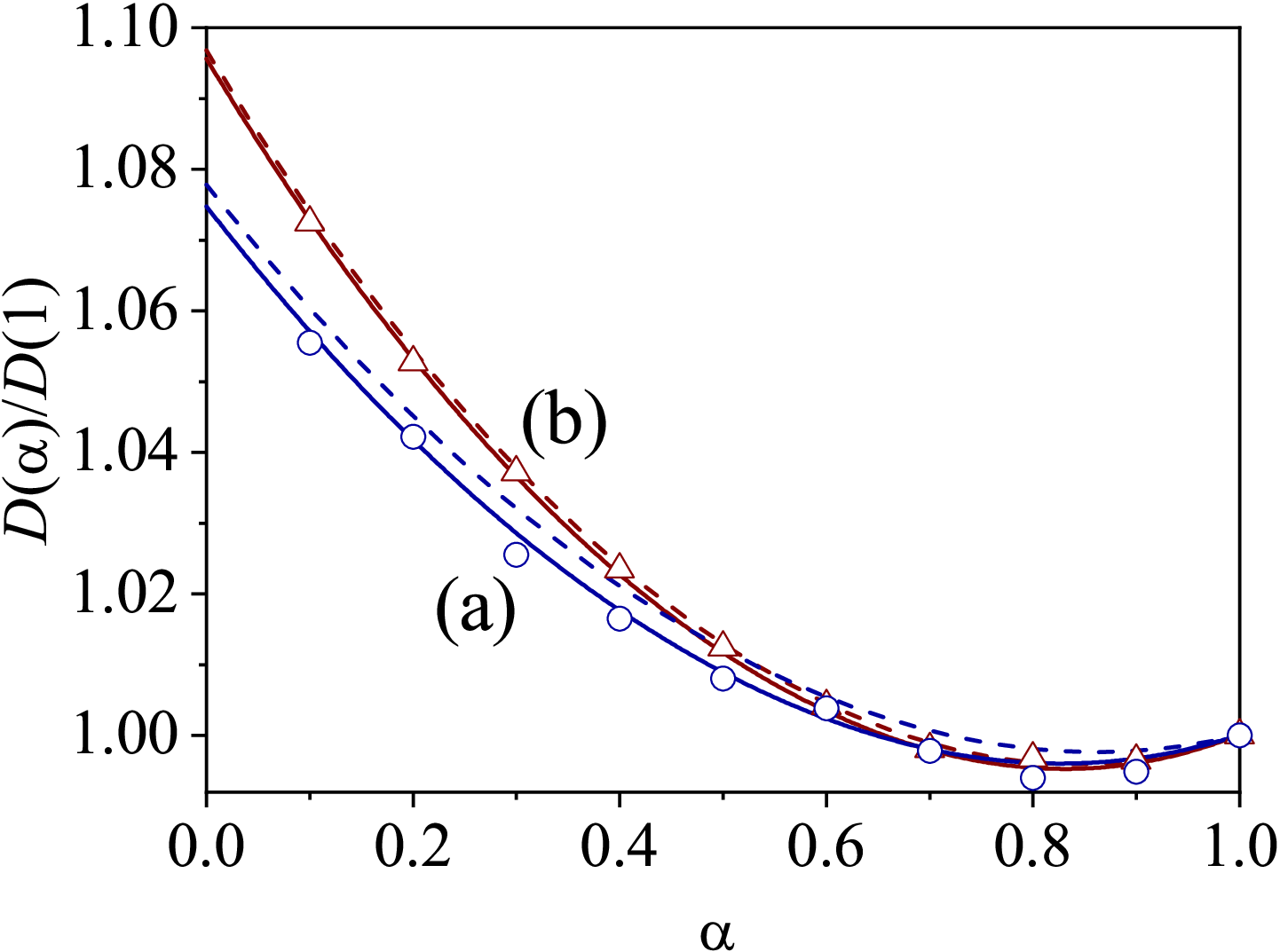}
\caption{Plot of the (reduced) diffusion coefficient $D(\al)/D(1)$ versus the (common) coefficient of restitution $\al$ for a three-dimensional ($d=3$) system  with $T_\text{b}^*$, $\phi=0.1$, and two different mixtures: (a) $m_0/m=0.5$ and $\sigma_0/\sigma=1$ (red line and triangles) and (b) $m_0/m=2$ and $\sigma_0/\sigma=1$ (blue line and circles). The symbols refer to the DSMC results.}
\label{fig5}
\end{figure}
\begin{figure}[h!]
\centering
\includegraphics[width=0.4\textwidth]{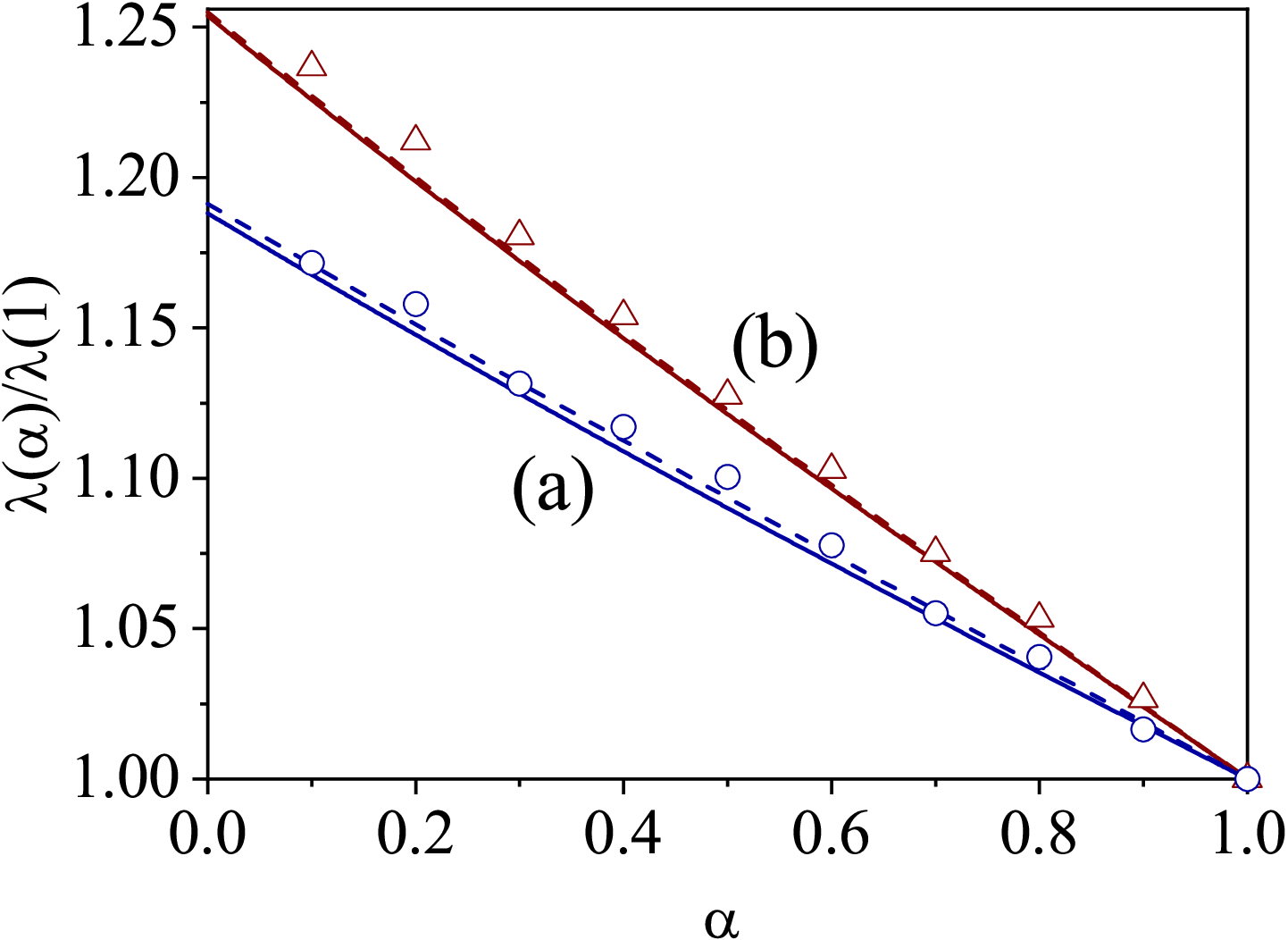}
\caption{Plot of the (reduced) mobility coefficient $\lambda(\al)/\lambda(1)$ versus the (common) coefficient of restitution $\al$ for a three-dimensional ($d=3$) system  with $T_\text{b}^*$, $\phi=0.1$, and two different mixtures: (a) $m_0/m=0.5$ and $\sigma_0/\sigma=1$ (red line and triangles) and (b) $m_0/m=2$ and $\sigma_0/\sigma=1$ (blue line and circles). The symbols refer to the DSMC results.}
\label{fig6}
\end{figure}

\subsection{DSMC simulations of $D$ and $\lambda$}

As in the case of the temperatures and the cumulants, to check the accuracy of the Sonine approximations we have solved the Enskog--Lorentz equation by means of the DSMC method described in section \ref{subsec1}. The diffusion $D$ and mobility $\lambda$ transport coefficients have been computed separately. 

Firstly, the calculation of the diffusion coefficient proceeds in the absence of any external field acting upon the intruder particles. In this scenario, Eq.\ \eqref{4.1} reads
\beq
\label{DSMC1}
 \frac{\partial n_0}{\partial t}=-\frac{D}{n_0}\nabla^2 n_0,
\eeq
where use has been made of the intruder particle flux equation \eqref{4.13} when $\mathbf{E}=\mathbf{0}$. In particular, the coefficient $D$ can be ascertained by evaluating the mean square displacement of the intruders \cite{M89}, as derived from the standard diffusion equation \eqref{DSMC1}. Specifically, we have
\beq
\label{DSMC2}
\frac{\partial}{\partial t}\langle |\mathbf{r}(t)-\mathbf{r}(0)|^2\rangle=2 d \frac{D_0}{n_0}.
\eeq
In this context, $\langle|\mathbf{r}(t)-\mathbf{r}(0)|\rangle$ is the ensemble-average distance traveled by the intruder up to the time $t$. 

On the other hand, the mobility of a tracer particle can be measured by applying a persistent yet small drag force $\mathbf{E} = E \mathbf{e}_x$ to the intruder particles. Over extended time intervals, the perturbed particles will reach a constant velocity $\lambda$, which is directly linked to the average distance travelled by the intruders by \cite{BBDLMP05}
\beq\label{DSMC3}
\langle(\mathbf{r}(t)-\mathbf{r}(0))\cdot\mathbf{e}_x\rangle\approx\lambda E t.
\eeq
Linearity of Eq.\ \eqref{DSMC3} has been checked in Ref.\ \cite{BBDLMP05} by changing the amplitude of the perturbation $E$.


Figures \ref{fig5} and \ref{fig6} show the dependence of the (reduced) transport coefficients $D(\al)/D(1)$ and $\lambda(\al)/\lambda(1)$ for a three-dimensional ($d=3$) system with $T_\text{b}^*$, $\phi=0.1$, and two different mixtures. Here, the diffusion $D$ and mobility $\lambda$ coefficients have been reduced with respect to their elastic limits $D(1)$ and $\lambda(1)$, respectively. Theoretical predictions given by the first and second Sonine approximations are compared with DSMC simulations. Although we observe that the first-Sonine approximation compares quite well with simulations, some small differences appear in the case of the diffusion coefficient for small mass ratios. These differences are mitigated by the second-Sonine approximation since it yields an excellent agreement with the DSMC results. Moreover, while the (reduced) mobility coefficient always increases with decreasing $\al$, a non-monotonic dependence of the (reduced) diffusion coefficient is present regardless the mass ratio considered. Figures \ref{fig5} and \ref{fig6} also highlight that the effect of the mass ratio on $\lambda(\al)/\lambda(1)$ is much more significant than for $D(\al)/D(1)$.

\subsection{Einstein relation}

\begin{figure}[h!]
\centering
\includegraphics[width=0.4\textwidth]{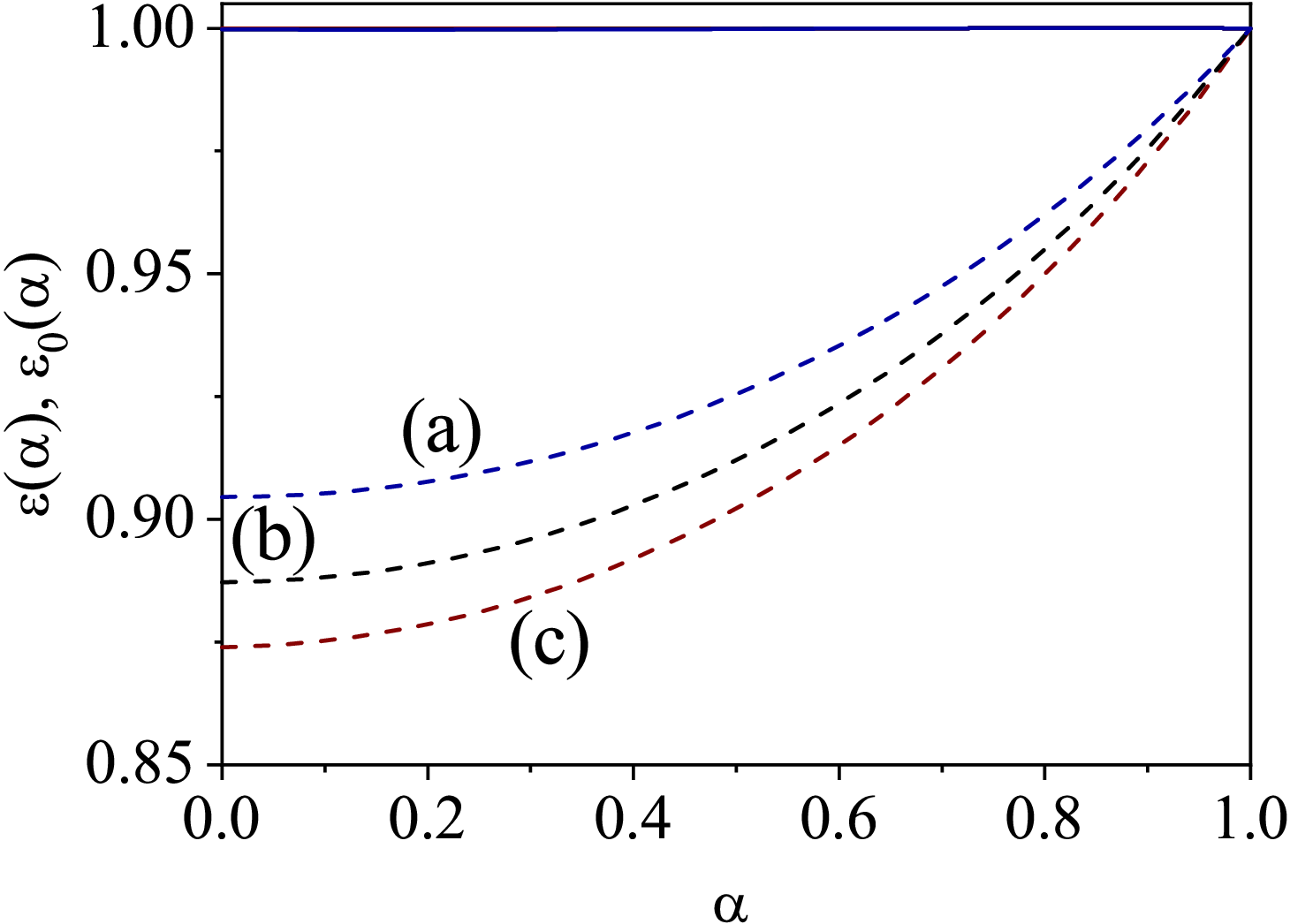}
\caption{Plot of the conventional $\epsilon$ (dashed lines)  and modified $\epsilon_0$ (solid lines) Einstein relation versus the (common) coefficient of restitution $\al$ for a three-dimensional ($d=3$) system  with $T_\text{b}^*$, $\phi=0.1$, and three different mixtures: (a) $m_0/m=0.5$ and $\sigma_0/\sigma=1$, (b) $m_0/m=1$ and $\sigma_0/\sigma=1$, and (c)  $m_0/m=2$ and $\sigma_0/\sigma=1$. The three lines corresponding to the modified Einstein relation are indistinguishable.}
\label{fig7}
\end{figure}

Once the transport coefficients are known, the conventional and modified relations can be explicitly obtained in terms of the parameters of the system up to the second-Sonine approximation. In the case of the conventional Einstein relation \eqref{4.17}, one gets the result
\beq
\label{4.30}
\epsilon[2]=\frac{D[2]}{T_\text{b}\lambda[2]}=T^*\frac{D^*[2]}{\lambda^*[2]}=T_0^*\Big(1-\frac{\nu_2^*}{\nu_4^*+3\gamma_0^*}c_0\Big),
\eeq
while in the case of the modified Einstein relation \eqref{4.18}, one achieves the expression 
\beq
\label{4.31}
\epsilon_0[2]=\frac{D[2]}{T_0\lambda[2]}=1-\frac{\nu_2^*}{\nu_4^*+3\gamma_0^*}c_0.
\eeq
It is quite apparent that while the conventional Einstein relation \eqref{4.30} fails due to both energy non-equipartition and non-Gaussian corrections to the distribution $f_0^{(0)}$, the departure of the modified Einstein relation \eqref{4.31} is only due to the latter feature ($c_0\neq 0$).  

To illustrate the dependence of both Einstein relations on the (common) coefficient of restitution $\al=\al_0$, Fig.\ \ref{fig7} shows $\epsilon$ and $\epsilon_0$ for several mixtures. While $\epsilon_0 \simeq 1$ for all the mixtures (in fact, the three curves practically collapse in a common curve in the scale of the vertical axis of Fig.\ \ref{fig7}), there are significant deviations from 1 in the conventional Einstein relation. In fact, the deviations of $\epsilon_0$ from 1 are much smaller than 1$\%$. This result contrasts with the ones previously derived for freely cooling \cite{DG01} and driven (with a Gaussian thermostat) \cite{G04} granular gases.   

In summary, the results derived here show no new surprises relative to the earlier work for dry granular gases \cite{G04,G08}: the origin of the brekdown of the modified Einstein relation is only due to the departure of the reference state from the Maxwell--Boltzmann distribution. However, in contrast to the previous works \cite{G04,G08}, the deviation of $\epsilon_0$ from 1 in granular suspensions is much smaller than the one found in driven granular gases, even for moderate densities and/or strong inelasticity.   

\section{Conclusions}
\label{sec5}

The main objective of this paper has been to analyze the validity of the conventional $\epsilon=D/T \lambda=1$ and modified $\epsilon_0=D/T_0 \lambda=1$ Einstein relations in a moderately dense granular suspension. The results have been derived in the framework of the (inelastic) Enskog kinetic equation, which applies to moderate densities. As usual in granular suspensions and due to the difficulties embodied in the description of systems constituted by two or more phases, a coarse-grained approach has been adopted. In this approach, the influence of the interstitial fluid on grains and intruders has been modeled through two different forces. Each one of the forces are composed by two terms: (i) a viscous drag term plus (ii) a stochastic Langevin-like term defined in terms of the background temperature $T_\text{b}$. Two different friction coefficients have been introduced introduced in the model; each one of them accounts for the interaction between the grains and intruders with the external bath. Thus, the starting kinetic equations for grains and intruders are the Enskog and the Enskog--Lorentz equations, respectively, with the addition of Fokker--Planck terms to each one of the above master equations. The present work extends previous studies performed by one of the authors of this paper in the case of driven \emph{dry} granular gases \cite{G04,G08}. 

To determine the explicit dependence of $\epsilon$ and $\epsilon_0$ on the parameter space of the system, the corresponding Enskog--Lorentz kinetic equation for intruders has been solved by means of the Chapman--Enskog method \cite{CC70} up to the first order in both the density gradient and the external field. As for molecular mixtures, the diffusion $D$ and mobility $\lambda$ transport coefficients are given in terms of a set of coupled linear integral equations, which are approximately solved by expanding the unknowns in a series of Sonine polynomials. Here, the series has been truncated by considering the two first relevant Sonine polynomials; this leads to the so-called first- and second-Sonine approximations to the coefficients $D$ and $\lambda$. The reliability of these theoretical results have been assessed via a comparison with computer simulations obtained by nuemrcially solving the Enskog--Lorentz equation by means of the DSMC method \cite{B94}. The comparison shows in general a good agreement between theory and simulations, specially in the case of the second-Sonine solution. This agreement provides a great confidence to the conclusions reached to the Einstein relations, which are based on approximate expressions for $D$ and $\lambda$. 

As expected from previous works \cite{G04,G08}, our results show that while the conventional Einstein relation is clearly violated, the deviations of the modified Einstein relation $\epsilon_0$ from 1 are very tiny. In particular, the deviations of $\epsilon_0$ from 1 are in general smaller than 1\% in the range of inelasticities and densities studied. This means that these deviations are much smaller than the ones reported in Ref.\ \cite{G08} for moderate densities, specially when the gas is driven by the Gaussian thermostat (see Fig.\ 6 of Ref.\ \cite{G08}). The fact that $\epsilon_0\simeq 1$ for granular suspensions is essentially due to the small magnitude of  the cumulant $c_0$, which is much smaller than in the driven case (compare for instance, Fig.\ 2 of Ref.\ \cite{G08} with Fig.\ \ref{fig4} of the present work). 

On other hand, the above conclusion disagrees with the computer simulation results obtained years ago by Puglisi \emph{et al.} \cite{PBV07}, which were subsequently confirmed in an experiment \cite{GPSV14}. In this experiment, the granular gas is driven by a shaker and the tracer is a rotating wheel immersed in the gas. The main claim in both papers is that the violation of the modified Einstein relation is mainly originated by the presence of spatial and velocity correlations which are relevant as density increases. Given that the Enskog equation takes into account the spatial correlations (through the pair correlation functions) \emph{but} neglects velocity correlation between the velocities of the particles which are about to collide (molecular chaos hypothesis), one could argue that the deviation of $\epsilon_0$ from 1 in the Enskog theory could be more important as both the density and inelasticity increase. However, our results indicate that the violation of the modified Einstein relation is still very small (and hence, undetectable in computer simulations) even when considers high densities and/or strong inelasticity. In this context and based on the Enskog results for granular suspensions at moderate densities, one could conclude that the origin of the deviation of $\epsilon_0$ from 1 is mainly due to velocity correlations, which are absent in the Enskog theory. These velocity correlations are expected to have a significant impact on $\epsilon_0$ for relatively high densities and/or high inelasticities.

In connection with the above point, one could include this sort of velocity correlations in the collision operator \cite{NEB98}. However, as mentioned in Ref.\ \cite{G08}, the inclusion of this new ingredient in the Enskog collision operator makes analytical calculations intractable since higher order correlations should be accounted for in the evaluation of the collision integrals. This type of calculation contrasts with the ones offered in this paper where the diffusion and mobility transport coefficients have been explicitly determined in terms of masses, diameters, coefficients of restitution, density and background temperature.      

Although some simulation computer works \cite{ML98,SM01,PTNE02} have clearly shown the failure of the molecular chaos assumption for inelastic collisions as the density increases, there is also some evidence in the granular literature on the usefulness of the Enskog theory for densities outside the dilute limit and inelasticities beyond the quasielastic limit. This evidence is supported by the agreement found at the level of the macroscopic properties between the Enskog results \cite{GD99a,L04,GDH07,GHD07} and those obtained from computer simulations \cite{LBD02,L04,DHGD02,MGAL06,LLC07} and real experiments \cite{YHCMW02,HYCMW04,BO07}. 

One of the limitations of the theoretical results presented in this paper is that they are approximated since they have been obtained by considering the second-Sonine approximation in the Chapman--Enskog solution. Exact results can be derived if one determines the coefficients $D$ and $\lambda$ by starting from the inelastic Maxwell model (IMM) for a dilute gas. As for molecular Maxwell gases \cite{CC70}, the collision rate of colliding particles in the IMM is assumed to be independent of the relative velocity. This simplification allows to express any moment of degree $k$ of the Boltzmann collisional operator in terms of velocity moments of degree $k$ or less than $k$ \cite{G19}. This feature of the Boltzmann collision operator of IMM opens the possibility of exactly determining the coefficients $D$ and $\lambda$. These results are presented in the Appendix \ref{appC}. According to these results, one concludes that the modified Einstein relation applies for IMM in any number of dimensions.

\acknowledgments

We acknowledge financial support from Grant PID2020-112936GB-I00 funded by MCIN/AEI/10.13039/501100011033, and  from Grant IB20079 funded by Junta de Extremadura (Spain) and by ERDF A way of making Europe. 

\appendix
\section{Expressions for the partial cooling  and the fourth-degree collisional moment}
\label{appA}

\begin{widetext}

In this Appendix we display the explicit expressions of the (reduced) partial cooling rate $\zeta_0^*$ and the fourth degree collisional moment $\Sigma_0$. Their forms are provided by Eqs.\ \eqref{3.9} and \eqref{3.12}, respectively, when nonlinear terms in $c_0$ and $c$ are neglected. The expressions of $\zeta_{00}$, $\zeta_{01}$, and $\zeta_{02}$ are given by \cite{GGG21}
\beq
\label{a1}
\zeta_{00}=\frac{2\sqrt{2}}{d}\left(\frac{\overline{\sigma}}{\sigma}\right)^{d-1}\frac{\chi_{0}}{\chi}\mu \left(\frac{1+\beta}{\beta}\right)^{1/2}(1+\alpha_{0})\left[1-\frac{1}{2}\mu (1+\alpha_{0})(1+\beta) \right],
\eeq
\beq
\label{a2}
\zeta_{01}=\frac{1}{2\sqrt{2}d}\left(\frac{\overline{\sigma}}{\sigma}\right)^{d-1}\frac{\chi_{0}}{\chi}\mu
\frac{(1+\beta)^{-3/2}}{\beta^{1/2}}(1+\alpha_{0}) \Big[3+4\beta-\frac{3}{2}\mu(1+\alpha_{0})(1+\beta) \Big],
\eeq
\beq
\label{a3}
\zeta_{02}=-\frac{1}{2\sqrt{2}d}\left(\frac{\overline{\sigma}}{\sigma}\right)^{d-1}\frac{\chi_{0}}{\chi}\mu
\left(\frac{1+\beta}{\beta}\right)^{-3/2}(1+\alpha_{0})
\Big[1+\frac{3}{2}\mu(1+\alpha_{0})(1+\beta) \Big].
\eeq
In the case of the fourth-degree collisional moment $\Sigma_0$, the expressions of $\Sigma_{00}$, $\Sigma_{01}$, and $\Sigma_{02}$ are \cite{GGG21}
\begin{eqnarray}
\label{a4}
\Sigma_{00} &=&\frac{1}{\sqrt{2}d(d+2)} \left(\frac{\overline{\sigma}}{\sigma}\right)^{d-1}\frac{\chi_{0}}{\chi}\mu
\left[ \beta(1+\beta)\right]^{-1/2}(1+\al_0)\Big\{-2\left[d+3+(d+2)\beta\right] +\mu\left( 1+\alpha_{0}\right) \left( 1+\beta \right)
\nonumber\\
& & \times
\left( 11+ d+\frac{d^2+5d+6}{d+3} \beta \right)-8\mu^{2}\left(1+\alpha_{0}\right)^{2}\left(1+\beta \right)^{2} +2\mu^{3}\left(
1+\alpha_{0}\right)^{3}\left( 1+\beta \right)^{3}\Big\},
\end{eqnarray}
\begin{eqnarray}
\label{a5}
\Sigma_{01} &=&
\frac{1}{8\sqrt{2}d(d+2)} \left(\frac{\overline{\sigma}}{\sigma}\right)^{d-1}\frac{\chi_{0}}{\chi}\mu \beta^{-1/2}\left(1+\beta\right)^{-5/2}
\left(1+\alpha_{0}\right)\Big\{-2\big[ 45+15d+(114+39d)\beta+(88+32d)\beta^{2}\nonumber\\
& & +(16+8d)\beta^{3}\big]+3\mu\left(1+\alpha_{0}\right) \left(1+\beta \right) \left[ 55+5d+9(10+d)\beta+4(8+d)\beta
^{2}\right]\nonumber\\
& & -24\mu^{2}\left(1+\alpha_{0}\right)^{2}\left(1+\beta \right)^{2}\left(5+4\beta
\right)+30\mu^{3}\left(1+\alpha_{0}\right)^{3}\left(1+\beta \right)^{3}\Big\},
\end{eqnarray}
\beqa
\label{a6}
\Sigma_{02}&=&\frac{1}{8\sqrt{2}d(d+2)} \left(\frac{\overline{\sigma}}{\sigma}\right)^{d-1}\frac{\chi_{0}}{\chi}\mu
\beta^{3/2}\left(1+\beta\right)^{-5/2} \left(
1+\alpha_{0}\right) \Big\{ 2\left[d-1+(d+2)\beta\right] +3\mu\left(1+\alpha_{0}\right) \left(1+\beta\right)\nonumber\\
& & \times \left[d-1+(d+2)\beta\right]
-24\mu^{2}\left( 1+\alpha_{0}\right)^{2}\left(1+\beta \right)^{2}+30\mu^{3}\left(1+\alpha_{0}\right)^{3}\left(1+\beta\right)^{3}\Big\} .
\eeqa

\end{widetext}

\section{Second Sonine approximation to the diffusion and mobility coefficients}
\label{appB}

Some details are provided in this Appendix in the calculation of the diffusion $D$ and mobility $\lambda$ coefficients up to the second Sonine approximation. Substitution of Eqs.\ \eqref{4.19} into the integral equations \eqref{4.11} and \eqref{4.12}, respectively, gives
\begin{widetext}
\beq
\label{b1}
\gamma_0\frac{\partial}{\partial\mathbf{v}}\cdot\mathbf{v} \Big(a_1 f_{0\text{M}}\mathbf{v}+a_2 f_{0\text{M}}\mathbf{S}_0\Big)
+\frac{\gamma_0 T_{\text{b}}}{m_0}\frac{\partial^2}{\partial v^2}\Big(a_1 f_{0\text{M}}\mathbf{v}+a_2 f_{0\text{M}}\mathbf{S}_0\Big)+a_1\chi_0 J_0^\text{B}[f_{0\text{M}}\mathbf{v},f]+a_2\chi_0 J_0^\text{B}[f_{0\text{M}}\mathbf{S}_0,f]=-\mathbf{v}f_0^{(0)},
\eeq
\beq
\label{b2}
\gamma_0\frac{\partial}{\partial\mathbf{v}}\cdot\mathbf{v} \Big(b_1 f_{0\text{M}}\mathbf{v}+b_2 f_{0\text{M}}\mathbf{S}_0\Big)
+\frac{\gamma_0 T_{\text{b}}}{m_0}\frac{\partial^2}{\partial v^2}\Big(b_1 f_{0\text{M}}\mathbf{v}+b_2 f_{0\text{M}}\mathbf{S}_0\Big)+b_1\chi_0 J_0^\text{B}[f_{0\text{M}}\mathbf{v},f]+
b_2\chi_0 J_0^\text{B}[f_{0\text{M}}\mathbf{S}_0,f]=-\frac{1}{m_0}
\frac{\partial}{\partial\mathbf{v}}f_0^{(0)}.
\eeq
\end{widetext}
Next, Eqs.\ \eqref{b1} and \eqref{b2} are multiplied by $\mathbf{v}$ and integrated over the velocity. The result is
\beq
\label{b3}
\left(\gamma_0+\nu_1\right)D+\frac{n_0 T_0^2}{m_0} \nu_2 a_2=\frac{n_0T_0}{m_0},
\eeq
\beq
\label{b4}
\left(\gamma_0+\nu_1\right)\lambda-\frac{n_0 T_0^2}{m_0} \nu_2 b_2=\frac{n_0}{m_0},
\eeq
where use has been made of the identities $a_1=(m_0 D/n_0 T_0)$, $b_1=-(m_0 \lambda/n_0T_0)$, and have introduced the quantities
\beq
\label{b5}
\nu_1=-\frac{m_0 \chi_0}{d n_0 T_0}\int d\mathbf{v}\; \mathbf{v}\cdot  J_0^\text{B}[f_{0\text{M}}\mathbf{v},f], \quad \nu_2=-\frac{m_0 \chi_0}{d n_0 T_0^2}\int d\mathbf{v}\; \mathbf{v}\cdot J_0^\text{B}[f_{0\text{M}}\mathbf{S}_0,f].
\eeq
If only the first Sonine correction is retained (i.e., $a_2=b_2=0$), then $D[1]=T_0 \lambda[1]$ and the modified Einstein relation \eqref{4.18} is verified.

To get the second-Sonine coefficients $a_2$ and $b_2$, one multiplies Eqs.\ \eqref{b1} and \eqref{b2} by $\mathbf{S}_0(\mathbf{v})$ and integrates over the velocity. The result is
\beq
\label{b7}
\frac{m_0}{n_0T_0^2}\Big[2\gamma_0\Big(1-\frac{T_\text{b}}{T_0}\Big)+\nu_3\Big]D+\left(3\gamma_0+\nu_4\right)a_2=\frac{c_0}{T_0},
\eeq
\beq
\label{b8}
-\frac{m_0}{n_0 T_0^2}\Big[2\gamma_0\Big(1-\frac{T_\text{b}}{T_0}\Big)+\nu_3\Big]\lambda+\left(3\gamma_0+\nu_4\right)b_2=0,
\eeq
where
\beq
\label{b9}
\nu_3=-\frac{2}{d(d+2)}\frac{m_0 \chi_0}{n_0 T_0^2}\int d\mathbf{v}\; \mathbf{S}_0\cdot J_0^\text{B}[f_{0\text{M}}\mathbf{v},f], \quad \nu_4=-\frac{2}{d(d+2)}\frac{m_0 \chi_0}{n_0 T_0^3}\int d\mathbf{v}\; \mathbf{S}_0\cdot  J_0^\text{B}[f_{0\text{M}}\mathbf{S}_0,f].
\eeq

In reduced units and by using matrix notation, Eqs.\ \eqref{b3} and \eqref{b7} along with Eqs.\ \eqref{b4} and \eqref{b8} can be rewritten as
\beq
\label{b11}
\left(
\begin{array}{cc}                                                                                           
\gamma_0^*+\nu_1^*&\tau_0^2 \nu_2^*\\
\frac{\nu_3^*+2\gamma_0^*\left(1-\frac{T_\text{b}^*}{T_0^*}\right)}{\tau_0^2}&3\gamma_0^*+\nu_4^*
\end{array}
\right)
\left(
\begin{array}{c}
D^*\\
a_2^*
\end{array}
\right)
=
\left(
\begin{array}{c}
\tau_0\\
\frac{c_0}{\tau_0}
\end{array}
\right),
\eeq
\beq
\label{b12}
\left(
\begin{array}{cc}
\gamma_0^*+\nu_1^*&-\tau_0^2 \nu_2^*\\
\nu_3^*+2\gamma_0^*\left(1-\frac{T_\text{b}^*}{T_0^*}\right)&-\tau_0^2 \left(3\gamma_0^*+\nu_4^*\right)
\end{array}
\right)
\left(
\begin{array}{c}
\lambda^*\\
b_2^*
\end{array}
\right)
=
\left(
\begin{array}{c}
1\\
0
\end{array}
\right).
\eeq
Here, $\nu_i^*=\nu_i/\nu$ ($i=1,\ldots,4$), $a_2^*=T\nu a_2$, and $b_2^*=T^2 \nu b_2$. From Eqs.\ \eqref{b11} and  \eqref{b12} one obtains the expressions \eqref{4.27} and \eqref{4.29} for the second-Sonine approximations to $D^*$ and $\lambda^*$, respectively.

The integrals involving the (reduced) collision frequencies $\nu_i^*$ have been computed in previous works \cite{GM07,GHD07,GV09} for a $d$-dimensional system when $f$ is replaced by the Mawellian distribution
\beq
\label{b13}
f_\text{M}(\mathbf{v})=n \left(\frac{m}{2\pi T}\right)^{d/2} \exp \left(-\frac{m v^2}{2 T}\right).
\eeq
In this case, the collision frequencies are given by
\begin{widetext}
\begin{equation}
\label{b14}
\nu_{1}^*=\frac{\sqrt{2}}{d}\left(\frac{\overline{\sigma}}{\sigma}
\right)^{d-1}\frac{\chi_0}{\chi}\mu (1+\alpha_0) \left(\frac{1+\beta}{\beta}\right)^{1/2}, \quad \nu_{2}^*=\frac{1}
{\sqrt{2}d}\left(\frac{\overline{\sigma}}{\sigma}\right)^{d-1}\frac{\chi_0}{\chi}\mu(1+\alpha_0)[\beta(1+\beta)]^{-1/2},
\end{equation}
\begin{equation}
\label{b16}
\nu_{3}^*=\frac{\sqrt{2}}
{d(d+2)}\left(\frac{\overline{\sigma}}{\sigma}\right)^{d-1}
\frac{\chi_0}{\chi}\mu(1+\alpha_0)\left(\frac{\beta}{1+\beta}\right)^{1/2}A_c,
\end{equation}
\begin{equation}
\label{b17}
\nu_{4}^*=\frac{1}
{\sqrt{2}d(d+2)}\left(\frac{\overline{\sigma}}{\sigma}\right)^{d-1}
\frac{\chi_0}{\chi}\mu(1+\alpha_0)\left(\frac{\beta}{1+\beta}\right)^{3/2}
\left[A_d-(d+2)\frac{1+\beta}{\beta} A_c\right],
\end{equation}
where
\begin{eqnarray}
\label{b18}
A_c&=& (d+2)(1+2\varpi)+\mu(1+\beta)\Big\{(d+2)(1-\alpha_0)
-[(11+d)\alpha_0-5d-7]\varpi\beta^{-1}\Big\}+3(d+3)\varpi^2\beta^{-1}\nonumber\\
& &+2\mu^2\left(2\alpha_0^{2}-\frac{d+3}{2}\alpha
_{12}+d+1\right)\beta^{-1}(1+\beta)^2- (d+2)\beta^{-1}(1+\beta),
\end{eqnarray}
\begin{eqnarray}
\label{b19}
A_d&=&2\mu^2\left(\frac{1+\beta}{\beta}\right)^{2}
\left(2\alpha_0^{2}-\frac{d+3}{2}\alpha_0+d+1\right)
\big[d+5+(d+2)\beta\big]-\mu(1+\beta) \Big\{\varpi\beta^{-2}[(d+5)+(d+2)\beta]
\nonumber\\
& & \times
[(11+d)\alpha_0
-5d-7]-\beta^{-1}[20+d(15-7\alpha_0)+d^2(1-\alpha_0)-28\alpha_0] -(d+2)^2(1-\alpha_0)\Big\}
\nonumber\\
& & +3(d+3)\varpi^2\beta^{-2}[d+5+(d+2)\beta]+ 2\varpi\beta^{-1}[24+11d+d^2+(d+2)^2\beta]
\nonumber\\
& & +(d+2)\beta^{-1} [d+3+(d+8)\beta]-(d+2)(1+\beta)\beta^{-2}
[d+3+(d+2)\beta].\nonumber\\
\end{eqnarray}
Here, $\varpi=(\mu_0/T_0)\left(T_0-T\right)$.
\end{widetext}

\section{Inelastic Maxwell model}
\label{appC}

In this Appendix we provide the exact results derived by considering the inelastic Maxwell model (IMM) for a dilute granular gas. The IMM is a further simplification of the inelastic hard sphere (IHS) model since it assumes that the collision rate of the colliding particles are independent of their relative velocity. In this model, the Boltzmann collision operator $J^{\text{IMM}}[f,f]$ of the granular gas reads \cite{G19} 
\beq
\label{c1}
J\left[\mathbf{v}_1|f,f\right]=\frac{\nu_\text{M}}{n S_d}  \int d\mathbf{v}_2\int d\widehat{\boldsymbol{\sigma}}
\Big[\alpha^{-1}f(\mathbf{v}_1'')f(\mathbf{v}_2'')-f(\mathbf{v}_1)f(\mathbf{v}_2)\Big],
\eeq
where $S_d=2\pi^{d/2}/\Gamma(d/2)$ is the total solid angle in $d$ dimensions and the velocities $\mathbf{v}_{1,2}''$ are related with $\mathbf{v}_{1,2}$ by Eqs.\ \eqref{2.3}. Moreover, $\nu_\text{M}$ is an effective collision frequency that is independent of velocity. This quantity can be seen as a free parameter of the model to be chosen to optimize the agreement with some proper quantity of interest obtained from the Boltzmann equation for IHS. In particular, if we chose $\nu_\text{M}$ to get the same expression of the cooling rate $\zeta$ as the one obtained in the Maxwellian approximation in the IHS model of diameter $\sigma$ [Eq.\ \eqref{2.18} with $c=0$], then one obtains the simple relationship $\nu_\text{M}=2\nu$ where $\nu$ is defined by Eq.\ \eqref{2.17} with $\chi=1$.

In the context of IMM, the Boltzmann--Lorentz collision operator $J_0^{\text{IMM}}[f_0,f]$ reads \cite{G19} 
\beq
\label{c2}
J_0^{\text{IMM}}[f_0,f]=\frac{\nu_{\text{M},0}}{n S_d}\int d\mathbf{v}_2\int d\widehat{\boldsymbol{\sigma}} \Big[\alpha_{0}^{-1}f_0(\mathbf{v}_1'')f(\mathbf{v}_2'')f_0(\mathbf{v}_1)f(\mathbf{v}_2)\Big],
\eeq
where $\nu_{\text{M},0}$ is an effective collision frequency for intruder-gas collisions and the relation between the velocities $\mathbf{v}_{1,2}''$ and $\mathbf{v}_{1,2}$ is given by Eqs.\ \eqref{2.22}. When the form \eqref{c2} of the operator $J_0^{\text{IMM}}[f_0,f]$ is substituted into the definition \eqref{3.3}, the partial cooling rate $\zeta_0 $ can be exactly determined for IMM. The result is \cite{G19}
\beq
\label{c3}
\zeta_{0}=\frac{2\nu_{\text{M},0}}{d}\mu (1+\alpha_{0})\left[1-\frac{1}{2}\mu (1+\alpha_{0})(1+\beta) \right].
\eeq    
Comparison of Eq.\ \eqref{c3} with Eq.\ \eqref{a1} (it gives $\zeta_0$ for IHS in the Maxwellian approximation, i.e., when $c=c_0=0$) yields the relation 
\beq
\label{c4}
\nu_{\text{M},0}=\sqrt{2}\left(\frac{\overline{\sigma}}{\sigma}\right)^{d-1}\left(\frac{1+\beta}{\beta}\right)^{1/2}\nu.
\eeq

The determination of the transport coefficients $D$ and $\lambda$ follows similar mathematical steps as those made in the case of IHS, except that the corresponding collision integrals appearing in the evaluation of these coefficients can be exactly computed. They are given by 
\beq
\label{c5}
\int d\mathbf{v}\; \mathbf{v}\cdot J_0^{\text{IMM}}[\boldsymbol{\mathcal{A}},f]=\nu_{\text{M},0} \mu (1+\alpha_0)D, \quad
\int d\mathbf{v}\; \mathbf{v}\cdot J_0^{\text{IMM}}[\boldsymbol{\mathcal{B}},f]=\nu_{\text{M},0} \mu (1+\alpha_0)\lambda.
\eeq
The final expressions of $D$ and $\lambda$ can be easily derived when one takes into account Eq.\ \eqref{c5}. The reults are
\beq
\label{c6}
D=\frac{n_0 T_0}{m_0}\left(\gamma_0+\nu_1\right)^{-1}, \quad \lambda=\frac{n_0}{m_0}\left(\gamma_0+\nu_1\right)^{-1},
\eeq
where 
\beq
\label{c7}
\nu_1=\mu (1+\alpha_0)\frac{\nu_{\text{M},0}}{d}=\frac{\sqrt{2}}{d}\left(\frac{\overline{\sigma}}{\sigma}\right)^{d-1}
\left(\frac{1+\beta}{\beta}\right)^{1/2}\mu (1+\alpha_0).
\eeq
Equation \eqref{c6} shows that the expressions of $D$ and $\lambda$ derived for IMM coincide with the ones obtained from the Boltzmann equation for IHS in the first-Sonine approximation when one neglects non-Gaussian corrections to the zeroth-order distributions ($c=c_0=0$). Thus, according to Eq.\ \eqref{c6}, $\epsilon_0=D/(T_0 \lambda)=1$ for IMM.

\bibliography{Brownian}

\end{document}